\documentclass[11pt]{article}
\usepackage{epsfig,amsfonts,amsthm,bm}
\usepackage{amsmath,amssymb}
\usepackage{cite}
\usepackage{xcolor}

\usepackage{float}

\renewcommand{\Re}{\mathrm{Re }}
\renewcommand{\Im}{\mathrm{Im }}
\newcommand{\doublet}[2]{ \left( \begin{array}{c}#1 \\ #2 \end{array}\right) }

\newcommand{\lr}[1]{ \langle #1 \rangle}

\newcommand{\mmmatrix}[9]{ \left(\!\! \begin{array}{ccc}#1 & #2 & #3\\ #4 & #5 & #6\\ #7 & #8 & #9\\ \end{array}\!\!\right) }

\newcommand{\toCP}{\xrightarrow{CP}}
\newcommand{\Br}{\mathrm{Br}}
\newcommand{\diag}{\mathrm{diag}}
\newcommand{\hsm}{h_{\mbox{\tiny SM}}}
\newcommand{\msm}{m_{\mbox{\tiny SM}}^2}
\newcommand{\SM}{\,\mbox{\tiny SM}}
\newcommand{\CKM}{\mbox{\tiny CKM}}
\newcommand{\exper}{\mbox{exp.}}
\newcommand{\LHC}{\mbox{\tiny LHC}}

\def\lsim{\mathrel{\rlap{\lower4pt\hbox{\hskip1pt$\sim$}}
    \raise1pt\hbox{$<$}}}         %less than or approx. symbol
\def\gsim{\mathrel{\rlap{\lower4pt\hbox{\hskip1pt$\sim$}}
    \raise1pt\hbox{$>$}}}         %greater than or approx. symbol

\title{Constraining the CP4-invariant three-Higgs-doublet model via top quark decays}

\author{Duanyang Zhao\thanks{E-mail: zhaody8@mail2.sysu.edu.cn},	
	Igor P. Ivanov\thanks{E-mail: ivanov@mail.sysu.edu.cn}
	\\
	{\small School of Physics and Astronomy, Sun Yat-sen University, 519082 Zhuhai, China}\\
	}

\begin{document}
\maketitle

\bigskip
\begin{abstract}
CP4 3HDM is a peculiar three-Higgs-doublet model in which a single symmetry leads to tight constraints
on the scalar and Yukawa sectors. In this models, tree-level flavor-changing neutral couplings are unavoidable;
however, as previously shown, their contributions to neutral meson oscillations can be suppressed.
Here, we explore the remaining quark flavor violating effects, which 
give rise to the top quark decays to light scalars, 
including the 125 GeV Higgs boson $\hsm$, as well as the magnitude of the $\hsm t\bar t$ coupling.
Utilizing the recently developed scanning procedure, in which observables are used as input,
we narrow down the viable options to a unique CP4-invariant Yukawa sector capable of
satisfying all meson oscillation and top quark constraints.
We present benchmark models that feature neutral or charged Higgs bosons 
lighter than the top quark, and we look forward to testing them further at the LHC 
and through flavor physics observables.
\end{abstract}

%\tableofcontents

\section{Introduction}\label{section-intro}

%\subsection{The Standard Model and N-Higgs Doublet Model}

There is a general consensus that the Standard Model (SM) cannot be the final word in our quest for the fundamental laws of microscopic physics but is a low-energy approximation to a hypothetical New Physics scenario yet to be established.
It is possible that the only manifestation of New Physics accessible to experiments at the electroweak energy scale is a non-minimal Brout-Englert-Higgs sector.
The 125 GeV Higgs boson observed in 2012 \cite{ATLAS:2012yve, CMS:2012qbp} and extensively studied by the ATLAS and CMS collaborations is compatible with the expected properties of the SM scalar particle. However, it can equally be just a member of a larger family of fundamental scalars.
The fact that this boson is very SM-like should not come as a surprise. 
Decades of experience in multi-Higgs model building \cite{Ivanov:2017dad} shows that by imposing global symmetries on the scalar and/or fermion sectors, one can naturally arrive at a situation where one Higgs boson is very SM-like, while the other scalars exhibit suppressed couplings to the $W$ and $Z$ bosons and the heaviest fermions, making them difficult to detect at the LHC. These new scalars can be rather light, with masses in the 100--200 GeV range, but if their production and decay channels are non-standard, they can escape detection. Interestingly, several hints of possible new scalars in this range have been detected by the LHC in the past few years and analyzed within various multi-Higgs scenarios \cite{Biekotter:2019kde,Benbrik:2022azi,Banik:2024ftv,Banik:2024ugs,Benbrik:2024ptw}.

Light new scalars can be searched for through their appearance in top quark decays such as 
\begin{equation}
	t \to H u_i, \quad u_i = u, c,\qquad
	t \to H^+ d_i, \quad d_i = d, s, b,\label{top} 
\end{equation}
followed by their subsequent decays into fermion pairs. An advantage of these processes is that one does not need to worry about potential suppression of their production channels, as top quark pairs are copiously produced at the LHC. Such decays are sensitive to the off-diagonal Yukawa couplings, offering a probe that complements the direct production of new scalars or flavor physics constraints. These signatures are predicted by various models, particularly multi-Higgs doublet models, see examples of such proposals in \cite{Aoki:2011wd,Akeroyd:2016ssd,Akeroyd:2018axd} along with a plethora of additional production and decay channels \cite{Arhrib:2021xmc,Hu:2022gwd,Cheung:2022ndq,Li:2023btx}. 
Interestingly, there remain benchmark points with relatively light Higgs bosons that pass the existing CMS and ATLAS searches
but may become excluded as more data is analyzed and additional search strategies are employed. Moreover, even if non-standard decays are hard to separate from the background processes, there always remains an integral observable, the total top quark decay width. Its Particle Data Group (PDG) 2024 value $\Gamma_{t}(\exper)=1.42^{+0.19}_{-0.15}$~GeV \cite{ParticleDataGroup:2024cfk} is compatible with the SM prediction $\Gamma_{t}^{\SM}=1.322$~GeV reported in \cite{Gao:2012ja}. These results limit the magnitude of the off-diagonal flavor couplings of any new Higgs boson lighter than the top quark,
including the SM-like Higgs $\hsm$.

In this paper, we explore how top quark decays constrain the CP4 3HDM, an intriguing version of the three-Higgs-doublet model based on an exact $CP$ symmetry of order 4. The model was proposed in \cite{Ivanov:2015mwl} and some of its phenomenological aspects were studied in \cite{Ferreira:2017tvy,Zhao:2023hws,Liu:2024aew}. A remarkable feature of this model is that imposing a single discrete symmetry leads to surprisingly tight connections between the scalar and fermion sectors. In this model, there is no natural flavor conservation, so that the tree-level flavor-changing neutral couplings (FCNC) are unavoidable, nor is there guaranteed scalar alignment. Still, it can easily accommodate the measured quark masses and mixing parameters, and with the aid of the scan procedure developed in \cite{Zhao:2023hws} one can potentially avoid the neutral meson oscillation constraints. Thus, CP4 3HDM still looks a viable multi-Higgs sector with an intriguing pattern of the off-diagonal fermion couplings. It is natural to check whether the versions of this model that feature light scalars can potentially survive the top-quark-based searches for new scalars.

The paper is organized as follows. In the next section, we briefly remind the reader of generalized $CP$ symmetries of higher orders and outline the main properties of the model CP4 3HDM. In section~\ref{section-couplings}, we write down the coupling matrices of the physical scalars with quarks and list the relevant expressions for decay widths. Next, in section~\ref{section-scan-setting}, we follow the procedures recently developed in \cite{Zhao:2023hws,Liu:2024aew} and describe an educated, physics-driven numerical scan of the scalar and Yukawa parameter spaces. In section~\ref{section-scan-results}, the results of this scan are compared with the experimental constraints from the LHC searches of the processes in Eq.~\eqref{top}. The lessons and conclusions emerging from this comparison are presented 
in section~\ref{section-conclusion}. 
Auxiliary expressions are collected in the Appendix.

Throughout the paper, the quark generations are labeled as $i$ or $j$, while the three Higgs doublets are marked with the subscripts $a$ or $b$. 
When dealing with neutral Higgs bosons, we switch from the three complex fields to the six real fields and label them with the indices $A$, $B$.

\section{CP4 3HDM: the present status of the model} \label{section-cp4}

\subsection{Order-4 $CP$ symmetry and the scalar potential}

To provide context, we begin with a brief sketch of the CP4 3HDM. 
Historically, multi-Higgs-doublet models---including the two-Higgs-doublet model (2HDM) and the three-Higgs-doublet model (3HDM)---emerged 
from the search for new opportunities for $CP$ violation, \cite{Lee:1973iz,Weinberg:1976hu,Branco:1999fs}.
It was later found that they could also accommodate new forms of $CP$ symmetry of higher order,
which are physically distinguishable from the traditional one.
In a model with several fields with identical gauge quantum numbers,
one can consider a general $CP$ transformation (GCP) that not only maps the fields to their conjugates
but also mixes them. A GCP acting on complex scalar fields $\phi_a$, $a = 1, \dots, N$,
can be written as \cite{Ecker:1987qp,Grimus:1995zi,Branco:1999fs,Weinberg:1995mt}
\begin{equation}
	\phi_a({\bf r}, t) \toCP {\cal CP}\,\phi_a({\bf r}, t)\, ({\cal CP})^{-1} = X_{ab}\phi_b^*(-{\bf r}, t), \quad X \in U(N)\,.
	\label{GCP}
\end{equation}
The conventional definition of $CP$ with $X_{ab} = \delta_{ab}$ is only one of many possible
choices; moreover, it is basis dependent.
If a model does not respect the conventional $CP$ but is invariant under a GCP with a suitable matrix $X$,
it is explicitly $CP$-conserving \cite{Branco:1999fs}.
Recall that the order $k$ of a transformation is defined by how many times one needs to apply it to arrive to the identity transformation.
The conventional $CP$ is of order 2, but a GCP, due to the presence of $X$, may be of a higher order.
In particular, the CP4 transformation is such a GCP of order 4.
Multi-Higgs-doublet models based on even higher order $CP$ symmetries are possible and were constructed in \cite{Ivanov:2018qni}.

Within the 2HDM, imposing CP4 on the scalar sector always leads to the usual $CP$ \cite{Ferreira:2009wh}.
In order to implement CP4 and avoid any conventional $CP$, one needs to pass to three Higgs doublets.
This model, dubbed CP4 3HDM, was constructed in \cite{Ivanov:2015mwl} building upon the results of \cite{Ivanov:2011ae}
and was found to possess remarkable features \cite{Ivanov:2015mwl,Aranda:2016qmp,Haber:2018iwr}.
In a suitable basis, the matrix $X$ of Eq.~\eqref{GCP} can be written as 
\begin{equation}
	X =  \left(\begin{array}{ccc}
		1 & 0 & 0 \\
		0 & 0 & i  \\
		0 & -i & 0
	\end{array}\right)\,.
	\label{CP4-def}
\end{equation}
Without loss of generality, the CP4-invariant 3HDM potential is $V=V_{0}+V_{1}$, where
%\begin{eqnarray}
%	V_0&=& - m_{11}^2 (\phi_1^\dagger \phi_1) - m_{22}^2 (\phi_2^\dagger \phi_2 + \phi_3^\dagger \phi_3) \nonumber\\
%	&&+ \lambda_1 (\phi_1^\dagger \phi_1)^2 + \lambda_2 \left[(\phi_2^\dagger \phi_2)^2 + (\phi_3^\dagger \phi_3)^2\right]
%	+ \lambda_{34} (\phi_1^\dagger \phi_1) (\phi_2^\dagger \phi_2 + \phi_3^\dagger \phi_3) \nonumber\\
%	&&- \lambda_4 \left[(\phi_1^\dagger \phi_1) (\phi_2^\dagger \phi_2) - (\phi_1^\dagger \phi_2)(\phi_2^\dagger \phi_1)
%	+ (\phi_1^\dagger \phi_1) (\phi_3^\dagger \phi_3) - (\phi_1^\dagger \phi_3)(\phi_3^\dagger \phi_1)\right]\nonumber\\
%	&&+
%	\lambda'_{34} (\phi_2^\dagger \phi_2) (\phi_3^\dagger \phi_3)
%	- \lambda'_4 \left[(\phi_2^\dagger \phi_2) (\phi_3^\dagger \phi_3) - (\phi_2^\dagger \phi_3)(\phi_3^\dagger \phi_2)\right]\,,
%	\label{V0}
%\end{eqnarray}
\begin{eqnarray}
	V_0&=& - m_{11}^2 (\phi_1^\dagger \phi_1) - m_{22}^2 (\phi_2^\dagger \phi_2 + \phi_3^\dagger \phi_3) 
	+ \lambda_1 (\phi_1^\dagger \phi_1)^2 + \lambda_2 \left[(\phi_2^\dagger \phi_2)^2 + (\phi_3^\dagger \phi_3)^2\right] \nonumber\\
	&& \hspace{-8mm}+\ \lambda_{34} (\phi_1^\dagger \phi_1) (\phi_2^\dagger \phi_2 + \phi_3^\dagger \phi_3) - \lambda_4 \left[(\phi_1^\dagger \phi_1) (\phi_2^\dagger \phi_2) - (\phi_1^\dagger \phi_2)(\phi_2^\dagger \phi_1)
	+ (\phi_1^\dagger \phi_1) (\phi_3^\dagger \phi_3) - (\phi_1^\dagger \phi_3)(\phi_3^\dagger \phi_1)\right]\nonumber\\
	&& \hspace{-8mm} +\ \lambda'_{34} (\phi_2^\dagger \phi_2) (\phi_3^\dagger \phi_3)
	- \lambda'_4 \left[(\phi_2^\dagger \phi_2) (\phi_3^\dagger \phi_3) - (\phi_2^\dagger \phi_3)(\phi_3^\dagger \phi_2)\right]
	\label{V0}
\end{eqnarray}
and
\begin{equation}
	V_1 = \lambda_5 (\phi_3^\dagger\phi_1)(\phi_2^\dagger\phi_1) +
	\lambda_8(\phi_2^\dagger \phi_3)^2 + \lambda_9(\phi_2^\dagger\phi_3)(\phi_2^\dagger\phi_2-\phi_3^\dagger\phi_3) + H.c.,
	\label{V1a}
\end{equation}
see details in Ref.~\cite{Ferreira:2017tvy}.
All parameters in $V_0$, as well as $\lambda_5$ in $V_1$, are real, while $\lambda_8$ and $\lambda_9$ must be complex
to avoid additional symmetries that would reduce the model to the $D_4$-invariant 3HDM
possessing the conventional $CP$ symmetry.
We assume that the parameters are chosen in such a way that the minimum of the potential is neutral \cite{Ferreira:2017tvy}.
We then parametrize the vacuum expectation values (vevs) $v_a = \sqrt{2}\lr{\phi_a^0}$ as
\begin{equation}
	(v_{1}, v_{2}, v_{3})= v(c_{\beta}, s_{\beta}c_{\psi}, s_{\beta}s_{\psi})\,,\label{vev}
\end{equation}
with the usual $v = 246$~GeV and the standard shorthand notation $c_x \equiv \cos(x)$ and $s_x \equiv \sin(x)$.
Note that the reality of the vevs is not an assumption but a consequence of the basis choice in which 
$s_{2\psi} \Im (\lambda_8) + c_{2\psi}\Im (\lambda_9) = 0$, see details in \cite{Liu:2024aew}. 

CP4 symmetry can be extended to the Yukawa sector 
\begin{equation}
	-{\cal L}_Y = \bar{Q}^0_L \Gamma_a \phi_a d_R^0 +
	\bar{Q}^0_L \Delta_a \tilde\phi_a  u_R^0 + H.c.,\label{Yukawa-general}
\end{equation}
leading to eight possible CP4-invariant patterns of the Yukawa matrices $(\Gamma_a, \Delta_a)$, \cite{Ferreira:2017tvy}.
%For completeness, these patterns are listed in Appendix~\ref{appendix-Yukawa}.
In Eq.~\eqref{Yukawa-general}, the quark generation indices are implicitly assumed, the superscript $0$ corresponds to the starting fields
before diagonalization of the mass matrices, and $\tilde\phi_a=i\sigma_{2}\phi^{\ast}_a$. 
By construction, CP4 mixes generations. Therefore, to avoid mass degenerate quarks, CP4 must be spontaneously broken,
which leads to the quark mass matrices and eventually to the masses of physical quarks and the rotation matrices.
It turns out that the Yukawa sector contains enough free parameters to accommodate
the experimentally measured quark masses and mixing angles as well as the appropriate amount of $CP$ violation.

\subsection{Phenomenology of CP4 3HDM: main features}

A peculiar feature of the CP4 3HDM is that a single symmetry leads to significant constraints on its free parameters
in the scalar and Yukawa sectors. 
In particular, it unavoidably leads to tree-level Higgs-mediated FCNCs, 
and it is not clear {\em \`a priori} whether these can be brought under control.

The first phenomenological study of the model reported in \cite{Ferreira:2017tvy} performed a numerical scan 
of the scalar and Yukawa parameter spaces and took into account the standard theoretical constraints on the scalar sector, 
electroweak precision observables, and the constraints coming from the kaon, $B$, and $B_s$-meson oscillation properties.
The $D$-meson constraints as well as non-standard top-quark couplings were not included in Ref.~\cite{Ferreira:2017tvy}. 
To suppress the FCNC induced by the SM-like Higgs boson,
the study assumed exact scalar alignment in the Higgs sector, an option that is possible but not compulsory.
Ref.~\cite{Ferreira:2017tvy} adopted the traditional scan procedure: one generates a random point in the vast parameter space,
computes the quark properties, which generically turn out to be way off from their experimental values, 
and then tries to optimize the global $\chi^2$ of the fit by an iterative procedure.
By construction, this scan is rather inefficient: the iterative procedure takes time, but the vast majority of the seed points 
still do not lead to any acceptable $\chi^2$. 
Nevertheless, the extensive scan of Ref.~\cite{Ferreira:2017tvy} yielded a few hundred points that seemed viable.

Almost all of these points contained one or two charged Higgs bosons lighter than the top quark,
so that new top decay channels open up, see Eq.~\eqref{top}.
In Ref.~\cite{Ivanov:2021pnr}, the parameter space points found in \cite{Ferreira:2017tvy} 
were tested against the LHC Run 2 searches for such top decays, and virtually all of them were ruled out 
because they either generated unacceptably large branching ratios 
$\Br(t \to H^+b)\times \Br(H^+\to c\bar s)$ or $\Br(t \to H^+b)\times \Br(H^+\to c\bar b)$
or led to the total top width exceeding its measured value.

After those two studies, it remained unclear whether the conflict between the scan results and the data represented 
a blow to the model or just exposed deficiencies in the scan procedures used in \cite{Ferreira:2017tvy}.
To get a better grasp on the model, a new scan should be run that takes into account 
the charm and top-quark constraints and, crucially, would be more efficient in identifying scenarios 
that are likely to satisfy experimental constraints.

This was the idea behind the two recent studies \cite{Zhao:2023hws,Liu:2024aew}, in which a 
physics-driven scan strategy was developed and applied to the Yukawa sector and the scalar sector of the model.
In a generic multi-Higgs-doublet model, one begins with Yukawa matrices in Eq.~\eqref{Yukawa-general}, 
multiplies them by the vevs of Eq.~\eqref{vev}, obtains the mass matrices, 
and diagonalizes them to get the fermion masses and mixing.
A peculiar feature of the CP4 3HDM is that this procedure can be inverted.
Namely, one can start with the physical quark masses and mixing parameters, choose suitable
quark rotation matrices that fit the CP4 patterns, and, for a generic set of vevs, 
recover the original Yukawa matrices.
This possibility, called in \cite{Zhao:2023hws} the inversion procedure, 
dramatically speeds up the scan in the Yukawa sector because, by construction, every seed point leads to physical quark properties.
Moreover, one can express the FCNC matrices in terms of the physical observables and quark rotation matrices,
which immediately gives some grasp on the magnitude of FCNC.
A key result of \cite{Zhao:2023hws} is that, out of eight possible CP4 invariant Yukawa scenarios, 
only two---labeled $(B_1,B_1)$ and $(A,B_2)$---had a chance to avoid unacceptably large contributions to the $K$, $B$, $B_s$, and $D$-meson oscillation parameters.
It is remarkable that this conclusion was reached without the need 
for a detailed scan in the scalar sector.

In \cite{Liu:2024aew}, an observable-driven inversion procedure was developed for the scalar sector of CP4 3HDM.
Instead of scanning the space of the scalar couplings $\lambda_i$ in Eqs.~\eqref{V0} and~\eqref{V1a},
a different set of variables was used, which allowed for better control over the phenomenological features of the model.
Namely, the proximity to the alignment limit and the admixture of various non-standard Yukawa structures
to the couplings of the SM-like Higgs boson $\hsm$ could be easily controlled.
Also, the regimes with light additional Higgs bosons could be separated from scenarios in which
all new scalars are heavier than the top quark. We draw the reader's attention to the fact that 
the CP4 3HDM does not possess the decoupling limit, which is a general result for multi-Higgs-doublet models
shaped by exact global symmetries that are spontaneously broken by the vevs \cite{Nebot:2019qvr,Carrolo:2021euy}.
As a result, perturbative unitarity constraints lead to upper bounds on the new scalar masses of the CP4 3HDM:
about 400~GeV for the lightest new Higgs boson and about 700~GeV for all additional scalars.

To summarize the present day situation with the CP4 3HDM phenomenology,
there are only two Yukawa scenarios that are promising, with quite distinct Yukawa patterns.
For the scalar sector, one can be either in the ``all heavy'' regime, 
in which all additional charged and neutral
Higgs bosons are heavier than the top quark, or allow for at least one non-standard Higgs boson, charged or neutral, 
to be lighter than the top quark.
In the former case, the top quark could still decay into the SM-like Higgs provided it possesses 
FCNC couplings in the up-quark sector. For this scenario, we wish to establish the magnitude 
of scalar misalignment still allowed by the present data.
In the latter case, our main concern is whether the new top decay channels in Eq.~\eqref{top}
together with meson oscillation properties rule out such a scenario altogether 
or leave room for benchmark points.

Before closing this section, we would like to address a natural question: 
why do we bother studying various sectors and sets of observables? 
Why not just perform an ``agnostic scan'' in the entire 
parameter space, couple its output to the standard computer codes such as \texttt{HiggsTools}, 
and find benchmark points that pass all constraints?
The answer is: experience shows that performing a scan in very multi-dimensional parameter space 
within default bounding boxes is inefficient in finding viable points of tightly constrained models.
Such a scan typically finds no viable points even if let run for days. 
However, guided by insights into the properties of specific sectors of a model's phenomenology, 
one can narrow down the scan, and the viable points begin to emerge. 
%We have seen this previously, and we observe the same phenomenon once again in the course of this study.
In parallel with this report, a general scan in the CP4 3HDM parameter space is being conducted, 
which relies on the results of \cite{Zhao:2023hws,Liu:2024aew} as well as of the present work.
The outcome of that study will be reported elsewhere.

\section{Top-Higgs couplings in the CP4 3HDM}\label{section-couplings}

\subsection{Parametrizing the scalar sector}

Description of the scalar sector and the patterns of Higgs-quark couplings become more transparent 
if we switch from the original doublets $\phi_a$ to a Higgs basis $\Phi_a$, in which only one doublet gets a non-zero vev:
$\lr{\Phi_1^0} = v/\sqrt{2}$, $\lr{\Phi_2} = \lr{\Phi_3} = 0$.
It is known that the choice of a Higgs basis is not unique in the 3HDM. 
In this work, we follow \cite{Zhao:2023hws,Liu:2024aew} and rely on the real vevs in Eq.~\eqref{vev}:
\begin{equation}
\left(\begin{array}{ccc}
\Phi_{1} \\
\Phi_{2} \\
\Phi_{3} \\
\end{array}\right)\
=\left(\begin{array}{ccc}
c_{\beta}&s_{\beta}c_{\psi}& s_{\beta}s_{\psi}\\
-s_{\beta}&c_{\beta}c_{\psi}& c_{\beta}s_{\psi}\\
0&-s_{\psi}& c_{\psi}\\
\end{array}\right)\
\left(\begin{array}{ccc}
\phi_{1} \\
\phi_{2} \\
\phi_{3} \\
\end{array}\right)\
\end{equation}
The three doublets can now be expanded as
\begin{equation}
	\Phi_1 = {1\over\sqrt{2}}\doublet{\sqrt{2}G^+}{v + \rho_1 + i G^0},\
	\Phi_2 = {1\over\sqrt{2}}\doublet{\sqrt{2}w_2^+}{\rho_2 + i \eta_2},\
	\Phi_3 = {1\over\sqrt{2}}\doublet{\sqrt{2}w_3^+}{\rho_3 + i \eta_3}.\label{expansion-Higgs-basis}
\end{equation}
%The transformation of the charged degrees of freedom is given by the same matrix ${\cal P}_{ij}$.
The mass matrix in the charged sector contains zeros for the would-be Goldstone boson $G^+$.
The remaining $2\times 2$ block is diagonalized in a straightforward way
and gives rise to the two physical charged Higgs bosons $H_1^\pm$ and $H_2^\pm$.
When dealing with the neutral fields, we arrange them as components of a six-dimensional real vector 
in the following way:
\begin{equation}
	\Phi_A = (G^0, \rho_1, \rho_2, \rho_3, \eta_3, \eta_2)\,.\label{Higgs-r}
\end{equation}
This particular ordering highlights yet another special structural feature of the CP4 3HDM: 
with this choice, the neutral scalar mass matrix takes a tridiagonal form, see details in \cite{Liu:2024aew}.
Once again, the would-be Goldstone mode $G^0$ decouples from the rest of the fields,
while the remaining $5\times 5$ block mixes the other components of $\Phi_A$.

Out of five physical neutral Higgs bosons, a special role is played by the SM-like boson $\hsm$,
which we identify with the detected 125 GeV scalar. We know from the LHC measurements
that it must be approximately aligned with the field $\rho_1$. Following \cite{Liu:2024aew},
we parametrize $\hsm$ with the four angles $\epsilon$, $\alpha$, $\gamma_1$, and $\gamma_2$: 
\begin{equation}
	\hsm =\rho_{1}\,c_{\epsilon}+s_{\epsilon} \left(c_{\alpha}c_{\gamma_{1}}\,\rho_{2}
	+c_{\alpha}s_{\gamma_{1}}\,\rho_{3}+s_{\alpha}c_{\gamma_{2}}\,\eta_{3}
	+s_{\alpha}s_{\gamma_{2}}\,\eta_{2}\right)\,.
	\label{hsm}
\end{equation}
The angle $\epsilon$ plays the role similar to the famous $\beta-\alpha$ of the 2HDM:
it controls the deviation of the model from the scalar alignment limit and, as a consequence, the $\hsm$ coupling to the $ZZ$ and $WW$ pairs.
Note that the first phenomenological study of CP4 3HDM assumed exact scalar alignment, $\epsilon = 0$.
In this work, we allow for a slight misalignment by using a non-zero $\epsilon$. 
It will be one of our tasks to determine how large $\epsilon$ can be without conflicting experiment. 

We scan the scalar parameter space in the manner similar to Ref.~\cite{Liu:2024aew} 
and adopt the following choice of the twelve free parameters:
%\begin{eqnarray}
%	\mbox{vev alignment:} && v\,, \quad \beta\,, \quad \psi\,, \nonumber\\
%	\mbox{SM-like Higgs properties:} && \msm\,, \quad \epsilon\,, \quad \alpha\,, \quad \gamma_1\,, \quad \gamma_2\,, \nonumber\\
%	\mbox{extra parameters:} && m_{H_1^\pm}^2\,, \quad m_{H_2^\pm}^2\,, \quad m_{11}^2 - m_{22}^2\,,\quad \lambda_{89}\,.  \label{input}
%\end{eqnarray}
\begin{equation}
	\{v\,, \  \beta\,, \  \psi\,,\  \msm\}\,,\quad \{\epsilon\,, \  \alpha\,, \  \gamma_1\,, \  \gamma_2\}\,,
	\quad \{m_{H_1^\pm}^2\,, \  m_{H_2^\pm}^2\,, \  m_{11}^2 - m_{22}^2\,,\  \lambda_{89}\}\,.\label{H-scan}
\end{equation}
Here, $v$ and $\msm$ are fixed by the measurements, while the ranges of the remaining parameters will be specified in the next section.
Scanning over these parameters instead of $\lambda_i$ gives us firm control over the properties of the SM-like Higgs boson.
All other observables in the scalar sector, including the masses and coupling patterns of the other four neutral Higgs bosons,
can be reconstructed from these parameters.

\subsection{Higgs-quark couplings}

As mentioned above, there exist eight possible CP4-invariant quark Yukawa sectors \cite{Ferreira:2017tvy}.
However, as shown in \cite{Zhao:2023hws}, only two of them avoid immediate conflict 
with the neural meson oscillation parameters.
The first one is labeled as case $(A, B_2)$, which means that the down-quark coupling matrices are of the following type:
	\begin{equation}
		\mbox{case $A$:}\quad \Gamma_1 = \mmmatrix{g_{11}}{g_{12}}{g_{13}}%
		{g_{12}^*}{g_{11}^*}{g_{13}^*}%
		{g_{31}}{g_{31}^*}{g_{33}}\,,\quad
		\Gamma_2 =0\,,\quad
		\Gamma_3 =0\,,\quad
		\label{caseA}
	\end{equation}
while the up-quark coupling matrices show a different pattern:
	\begin{equation}
		\mbox{case $B_2$:}\quad \Delta_1 = \mmmatrix{0}{0}{d_{13}}{0}{0}{d_{13}^*}{0}{0}{d_{33}}\,,\quad
		\Delta_2 = \mmmatrix{d_{11}}{d_{12}}{0}{d_{21}}{d_{22}}{0}{d_{31}}{d_{32}}{0}\,,\quad
		\Delta_3 =  \mmmatrix{d_{22}^*}{-d_{21}^*}{0}{d_{12}^*}{-d_{11}^*}{0}{d_{32}^*}{-d_{31}^*}{0}\,.
		\label{caseB2}
	\end{equation}
In these matrices, all entries apart from the element $(33)$ can be complex.
Note also tight correlations between $\Delta_2$ and $\Delta_3$.
The second CP4-invariant Yukawa structure that was considered potentially viable in \cite{Zhao:2023hws} 
is labeled as $(B_1, B_1)$. Its down-quark and up-quark Yukawa matrices
can be obtained from \eqref{caseB2} by transposition.
We do not show these matrices explicitly because, as we will report in the next section,
this scenario will be ruled out by the neutral meson oscillation constraints.
Thus, $(A, B_2)$ will be our main scenario, the only one that survives all the flavor constraints explored in this paper.

Next, we write the quark Yukawa sector in the Higgs basis as
\begin{eqnarray}
-{\cal L}_Y  =\frac{\sqrt{2}}{v} \bar{Q}^0_L \left(\Phi_1 M_d^0 + \Phi_2 N_{d2}^0 + \Phi_3 N_{d3}^0\right) d_R^0
 +\frac{\sqrt{2}}{v}\bar{Q}^0_L  \left(\tilde\Phi_1 M_u^0 + \tilde\Phi_2 N_{u2}^0 + \tilde\Phi_3 N_{u3}^0\right) u_R^0 + H.c. \label{L-Higgs-basis}
\end{eqnarray}
Here, 
\begin{eqnarray}
N_{d2}^0 = - M_d^0 \tan\beta + \frac{v}{\sqrt{2}c_\beta}(\Gamma_2 c_\psi + \Gamma_3 s_\psi)\,, \quad
N_{d3}^0 = \frac{v}{\sqrt{2}}(-\Gamma_2 s_\psi + \Gamma_3 c_\psi)\,,\label{Nd20Nd30-general}
\end{eqnarray}
and similarly
\begin{eqnarray}
N_{u2}^0 =- M_u^0 \tan\beta + \frac{v}{\sqrt{2}c_\beta}(\Delta_2 c_\psi + \Delta_3 s_\psi)\,, \quad
N_{u3}^0 = \frac{v}{\sqrt{2}}(-\Delta_2 s_\psi + \Delta_3 c_\psi)\,.
\end{eqnarray}
We remind the reader that the superscript "0" indicates the original bases in the quark field spaces. 
These need to be rotated as 
$d_L^0 = V_{dL} d_L$, $d_R^0 = V_{dR}d_R$, $u_L^0 = V_{uL} u_L$, $u_R^0 = V_{uR} u_R$
%\begin{equation}
%	d_L^0 = V_{dL} d_L\,, \quad d_R^0 = V_{dR} d_R\,, \quad
%	u_L^0 = V_{uL} u_L\,, \quad u_R^0 = V_{uR} u_R\,.\label{quark-rotations}
%\end{equation}
to make the quark mass matrices diagonal: 
$M_d = V_{dL}^\dagger M_d^0 V_{dR} = {\rm diag}(m_d, m_s, m_b)$ and 
$M_u = V_{uL}^\dagger M_u^0 V_{uR} = {\rm diag}(m_u, m_c, m_t)$.
The mismatch between the rotation matrices in the left-handed up and down-quark spaces
gives rise to the Cabibbo-Kobayashi-Maskawa (CKM) matrix $V_{\CKM} = V_{uL}^\dagger V_{dL}$.
The coupling patterns of the Higgs basis doublets $\Phi_2$ and $\Phi_3$
with the physical quarks are described by the matrices
\begin{equation}
N_{d2} = V_{dL}^\dagger N_{d2}^0 V_{dR}\,, \quad N_{d3} = V_{dL}^\dagger N_{d3}^0 V_{dR}\,,\quad
N_{u2} = V_{uL}^\dagger N_{u2}^0 V_{uR}\,, \quad N_{u3} = V_{uL}^\dagger N_{u3}^0 V_{uR}\,.\label{FCNCmatrices}
\end{equation}
Explicit expressions for the specific cases $(A, B_2)$ and $(B_1, B_1)$ can be found in \cite{Zhao:2023hws}.
In particular, for the case $A$ in the down-quark sector, $N_{d2} = -M_d \tan\beta$ and $N_{d3} =0$.
%This means that, even without FCNC in the down-quark-Higgs couplings, the diagonal couplings 
%of the SM-like Higgs such as $\hsm b\bar b$ can differ from the SM value if the misalignment angle $\epsilon$
%or the value of $\tan\beta$ are not suppressed.

In the above expressions, the quark fields are already mass eigenstates
while the scalar fields are still defined in the Higgs basis and must be further rotated into the scalar mass eigenstates.
Let us begin with the neutral scalars.
The neutral boson mass matrix ${\cal M}_n$ was explicitly given in the Higgs basis in \cite{Liu:2024aew}
as a $6\times 6$ real symmetric matrix in the space of the real components $\Phi_A$, see Eq.~\eqref{Higgs-r}.
We take the liberty to counts them as $A = 0, 1, \dots 5$, reserving $0$ for the neutral would-be Goldstone $G^0$.
Next, let us denote by $S_{AB}$ the orthogonal $6\times 6$ matrix that diagonalizes ${\cal M}_{n}$:
\begin{equation}
	S^{T}\, {\cal M}_{n}\, S = \diag\left(0, m_{H_{1}}^{2}, m_{H_{2}}^{2}, m_{H_{3}}^{2}, m_{H_{4}}^{2}, m_{H_{5}}^{2}\right)\,,
	\quad m_{H_A} \le m_{H_{A+1}}\,.
	\label{neutral-Higgs-diagonal}
\end{equation}
This matrix links the Higgs basis fields $\Phi_A$ with neutral boson eigenstates: $\Phi_A = S_{AB} H_B$,
where $H_A=(G^{0}, H_{1}, H_{2}, H_{3}, H_{4}, H_{5})$.
Since $G^0$ decouples, the matrix $S_{AB}$ has $S_{00} = 1$, $S_{0B} = 0$, and contains a non-trivial $5\times 5$ block
for $A, B = 1, \dots, 5$.
With this notation, we can write the interaction lagrangian of the five physical neutral bosons with quark pairs as
\begin{eqnarray}
	-{\cal L}_{n}
	=\frac{1}{v} \left( \overline{d}_{L}\widetilde{N}_{dA}d_{R} + 
	\overline{u}_{L}\widetilde{N}_{uA}u_{R}\right)H_{A}+H.c.,\label{L-n-p}
\end{eqnarray}
where the quark coupling matrices for the Higgs boson $H_A$ are given by
\begin{eqnarray}
	\widetilde{N}_{dA}&=&M_{d}S_{1A} + N_{d2}(S_{2A}+iS_{5A})+N_{d3}(S_{3A}+iS_{4A})\,,\nonumber\\
	\widetilde{N}_{uA}&=&M_{u}S_{1A} + N_{u2}(S_{2A}-iS_{5A})+N_{u3}(S_{3A}-iS_{4A})\,.\label{L-n-c}
\end{eqnarray}
One among the five $H_A$ bosons is the SM-like Higgs bosons $\hsm$; {\em \`a priori}, it can occupy any position in the spectrum. 
With the parameterization \eqref{hsm}, we can write its couplings to quark pairs as above 
with the following coefficients:
\begin{eqnarray}
	S_{1\SM}=c_{\epsilon}\,,\quad S_{2\SM}=s_{\epsilon}c_{\alpha}c_{\gamma_{1}}\,,\quad
	S_{3\SM}=s_{\epsilon}c_{\alpha}s_{\gamma_{1}}\,,\quad S_{4\SM}=s_{\epsilon}s_{\alpha}c_{\gamma_{2}}\,,\quad
	S_{5\SM}=s_{\epsilon}s_{\alpha}s_{\gamma_{2}}\,.\label{L-n-p2}
\end{eqnarray}

Turning now to the charged Higgs bosons, we find that their couplings inherit the neutral matrix structure.
In the Higgs basis, with the charged components of the doublets \eqref{expansion-Higgs-basis} labeled as $(G^+, w_2^+, w_3^+)$, 
we obtain
\begin{equation}
	-{\cal L}_{ch} = \frac{\sqrt{2}}{v}\overline{u}_L^0 (N_{d2}^0 w_2^+ + N_{d3}^0 w_3^+) d_R^0
	-\frac{\sqrt{2}}{v}\overline{d}_L^{\,0} (N_{u2}^0 w_2^- + N_{u3}^0 w_3^-) u_R^0 + H.c.
\end{equation}
After the quark field rotation, we arrive at
\begin{equation}
	-{\cal L}_{ch} = \frac{\sqrt{2}}{v}\left(\overline{u}_L V_{\CKM} N_{da} d_R - \overline{u}_R N_{ua}^\dagger V_{\CKM} d_L\right) w_a^+ 
		+ H.c.\label{charged-couplings}
\end{equation}
Thus, once all the matrices $N$ are known, we can also directly establish how the physical quarks couple
to the charged Higgs fields. Let us denote by ${\cal M}_{ch}$ the charged Higgs boson mass matrix in the Higgs basis. 
It is a $2\times 2$ real symmetric matrix, for an explicit expression, see \cite{Liu:2024aew}.
Let $R_{ab}$ be the $2\times 2$ orthogonal matrix that diagonalizes ${\cal M}_{ch}$:
\begin{equation}
	R^{T}{\cal M}_{ch}R=\diag\left(m_{H_1^+}^{2}, m_{H_2^+}^{2}\right)\,, \quad m_{H_1^+} < m_{H_2^+}\,.\label{charged-Higgs-diagonal}
\end{equation}
Then, the physical charged Higgs bosons couple with the physical quarks according to the following lagrangian:
%\begin{equation}
%	-{\cal L}_{ch} = \frac{\sqrt{2}}{v}\left(\overline{u}_L V_{\CKM} N_{da} d_R - \overline{u}_R N_{ua}^\dagger V_{\CKM} d_L\right) R_{ab}H_b^+ 
%	+ H.c.\label{charged-couplings-2}
%\end{equation}
\begin{equation}
	-{\cal L}_{ch} = \left(\overline{u}_L A_b \,d_R - \overline{u}_R B_b^\dagger \, d_L\right) H_b^+ 
	+ H.c.,\label{charged-couplings-2}
\end{equation}
where the matrices $A$ and $B$ are
\begin{equation}
	A_{b,ij}=\frac{\sqrt{2}}{v} \, \left(V_{\CKM}N_{da}\right)_{ij} R_{ab}\,,\quad
	B^{\dagger}_{b,ij}=\frac{\sqrt{2}}{v} \left(N_{ua}^{\dagger} V_{\CKM}\right)_{ij} R_{ab}\,,\label{matrices-A-B}
\end{equation} 
where $a=2,3$ and $b=1,2$.
When parametrizing the Yukawa sector, we follow the inversion procedure developed in \cite{Zhao:2023hws}.
We take the physical quark masses and CKM matrix as input parameters,
complement them with suitably chosen right-handed field rotation matrices $V_{dR}$ and $V_{uR}$,
and reconstruct the mass matrices $M_d^0$ and $M_u^0$ before diagonalization.
Next, by using the vev parameters $v$, $\beta$, $\psi$ as input from the scalar sector,
we uniquely determine the individual matrices $\Gamma_a$ and $\Delta_a$, as well as the FCNC matrices
$N_{d2}$, $N_{d3}$ and $N_{u2}$, $N_{u3}$. We stress that this inversion is not possible in a general multi-Higgs-doublet model.
Its existence in CP4 3HDM is non-trivial and represents a convenient feature of the model.

\subsection{Non-standard top quark decays}\label{subsection-top-decays}

With these explicit form for the Higgs-quark couplings,
we can now give expressions for the non-standard tree level decays of the top quark
into light Higgs bosons.
Each light charged Higgs boson $H^+_b$, $b=1,2$, gives the following contribution to the top quark decay width: 
\begin{equation}
	\Gamma_{t\rightarrow d_{j}H_{b}^{+}}=
	\frac{\sqrt{\lambda(m_{t},m_{j},m_{H})}}{32\pi m_{t}^{3}}
	\left[\left(|A_{b, 3j}|^{2}+|B^{\dagger}_{b, 3j}|^{2}\right)(m_{t}^{2}+m_{j}^{2}-m_{H}^{2})
	- 4m_{j}m_{t}\Re(A_{b, 3j}B_{b, j3})\right]\,.
	\label{decay-1}
\end{equation}
Here, $\lambda(m_{1},m_{2},m_{3})=m_{1}^{4}+m_{2}^{4}+m_{3}^{4}-2m_{1}^{2}m_{2}^{2}
-2m_{2}^{2}m_{3}^{2}-2m_{1}^{2}m_{3}^{2}$; we also used the shorthand notation $m_j \equiv m_{d_j}$
and $m_H \equiv m_{H_b^+}$. 
The charged Higgs boson can then decay to a quark pair with the partial decay width
\begin{equation}
\Gamma_{H_{b}^{+}\rightarrow u_{i}\overline{d}_{j}}=
\frac{N_{c}\sqrt{\lambda(m_{H},m_{i},m_{j})}}{16\pi m_{H}^{3}}
\left[\left(|A_{b, ij}|^{2}+|B^{\dagger}_{b, ij}|^{2}\right)(m_{H}^{2}-m_{i}^{2}-m_{j}^{2})
+ 4m_{i}m_{j}\Re(A_{b, ij}B_{b, ji})\right]\,.
\label{decay-2}
\end{equation}
We checked that these expressions are consistent with those in \cite{Ivanov:2021pnr} if we ignore the light quark masses
and take the alignment limit.

Each light neutral Higgs boson $H_B$, $B=1,\dots, 5$, gives the following contribution to the top quark decay width: 
\begin{equation}
\Gamma_{t\rightarrow u_{i}H_{B}}=\frac{\sqrt{\lambda(m_{t},m_{i},m_{H})}}{32\pi m_{t}^{3}v^{2}}
\left[\left(|\widetilde{N}_{uB, i3}|^{2}+|\widetilde{N}^{\dagger}_{uB, i3}|^{2}\right)
(m_{t}^{2}+m_{i}^{2}-m_{H}^{2})+4m_{t}m_{i}\Re(\widetilde{N}_{uB, i3}\widetilde{N}_{uB, 3i})\right]\,,
\label{decay-3}
\end{equation}
with the matrices $\widetilde{N}$ defined in Eq.~\eqref{L-n-c}.
Its partial decay width to each up-quark pair is given by 
\begin{equation}
\Gamma_{H_{B}\rightarrow u_{i}\overline{u}_{j}}=
\frac{N_{c}\sqrt{\lambda(m_{H},m_{i},m_{j})}}{16\pi m_{H}^{3}v^{2}}
\left[\left(|\widetilde{N}_{uB, ij}|^{2}+|\widetilde{N}^{\dagger}_{uB, ij}|^{2}\right)
(m_{H}^{2}-m_{i}^{2}-m_{j}^{2})-4m_{i}m_{j}\Re(\widetilde{N}_{uB, ij}\widetilde{N}_{uB, ji})\right]\,.\label{decay-4}
\end{equation}
For the decay to down-quark pairs, $\Gamma_{H_{B}\rightarrow d_{i}\overline{d}_{j}}$ has same form with 
$\widetilde{N}_{uB}$ replaced by $\widetilde{N}_{dB}$.

\section{Numerical study: the scan procedure} 
\label{section-scan-setting}

\subsection{The top decays used}

Our main focus is on the top quark decays to light Higgs bosons and the constraints on the CP4 3HDM parameters 
that follow from a comparison of the model predictions with experimental data.
%Specifically, we include three types of experimental results: the LHC searches for the top decays
%to charged Higgs bosons, $t \to H^+ d_i$, the LHC searches for the top decays to neutral Higgs bosons,
%$t \to H u_i$, where $H$ could be either the SM-like or a new scalar, and the measurements of the total top quark decay width,
%which could be enhanced with respect to the SM value due to the presence of new channels.
For each charged Higgs boson $H^+_a$ lighter than the top-quark, we compute the decay width $t\rightarrow d_{i}H_{a}^{+}$ 
and all the partial decay widths $H_{a}^{+}\rightarrow u_{j}\overline{d}_{i}$. 
In principle, charged Higgs bosons can also decay to gauge bosons or other scalars $H_{a}^{+}\rightarrow W^{+}H_{B}$, 
$H_{a}^{+}\rightarrow ZH_{a'}^{+}$, and $H_{a}^{+}\rightarrow H_{B}H_{a'}^{+}$. 
However, these are expected to be suppressed by the very limited phase space, and we do not take them into account.

To compare the model with the LHC data, we follow the procedure of \cite{Ivanov:2021pnr} and 
define the products of branching ratios:
\begin{equation}
	P_{cb}=\Br(t\rightarrow bH_{a}^{+})\times \Br(H_{a}^{+}\rightarrow c\overline{b})\,,\qquad				%% previous A1
	P_{cs}=\Br(t\rightarrow bH_{a}^{+})\times \Br(H_{a}^{+}\rightarrow c\overline{s})\,.\label{br1}			%% previous A2	
\end{equation}
Both quantities have been probed by the CMS and ATLAS searches,
see Refs.~\cite{CMS:2018dzl,ATLAS:2023bzb} for the $c\overline{b}$ final state
and Refs.~\cite{ATLAS:2013uxj,CMS:2015yvc,CMS:2020osd,ATLAS:2024oqu} for the $c\overline{s}$ pair.
In this work, we use the strongest experimental upper bound $P_{cb}(\exper)$ that comes from the ATLAS 2023 study \cite{ATLAS:2023bzb} 
and varies between $0.15\%$ and $0.42\%$ for the charged Higgs boson mass from 60 to 160 GeV,
as well as the strongest upper bound $P_{cs}(\exper)$ from the ATLAS 2024 study \cite{ATLAS:2024oqu} 
that varies between $0.066\%$ and $3.6\%$ in the 60 to 168 GeV mass range. 
%The complete relation between upper limit and mass of charge Higgs could be found in Figure 8 of ref\cite{ATLAS:2023bzb} and Figure 6 of ref\cite{ATLAS:2024oqu}. 

When performing a scan, we can generate examples of the CP4 3HDM model with various charged Higgs boson masses.
Since the upper limits depend on the mass, we compare the predictions of the model with the experimental limits
by calculating the ratios
\begin{equation}
	R_{cb} = \frac{P_{cb}}{P_{cb}(\exper)}\,, \qquad 
	R_{cs} = \frac{P_{cs}}{P_{cs}(\exper)}\,, \label{R-cbcs}
\end{equation}
where $P_{cb}(\exper)$ and $P_{cs}(\exper)$ are taken for the value of the charged Higgs boson mass.
If a benchmark point contains two light charged Higgs bosons $H_a^+$,
we compute these ratios for both of them. 
%, select the largest $R_{cb}$ and the largest $R_{cs}$ of the two,
%and require them to stay below one.

For the neutral Higgs bosons lighter than the top, we define similar ratios:
\begin{equation}
	P_u = \Br(t\rightarrow uH_A)\times \Br(H_A\rightarrow b\overline{b})\,, \qquad						%% previous A3
	P_c = \Br(t\rightarrow cH_A)\times \Br(H_A\rightarrow b\overline{b})\,.								%% previous A4
	\label{br2}
\end{equation}
This time, for the correct computation of the branching ratios $H_A\rightarrow b\overline{b}$, 
we must take into account not only all $q'\overline{q}$ final states but also $gg$, 
$WW^*$, and $ZZ^*$ decays, which can be especially important
for $m_{H_A}$ close to the upper end of the mass interval. In appendix~\ref{appendix-nonqq},
we provide details on how these were computed.
Suppressed decay channels such as $H_A\rightarrow \gamma\gamma$, 
$W^{\pm}H_{a}^{\mp}$, $ZH_B$, and $H_{a}^{+}H_{b}^{-}$
were ignored in the total $H_A$ decay width calculation.

For $P_u(\exper)$, we use the upper bounds reported by ATLAS \cite{ATLAS:2023mcc}, which vary from $0.019\%$ to $0.062\%$ 
in the additional neutral scalar mass range 20 to 160 GeV. The upper limits on $P_c(\exper)$ are also taken from \cite{ATLAS:2023mcc} 
and vary between $0.018\%$ and $0.078\%$. 
Similarly to Eq.~\eqref{R-cbcs}, we define the ratios
\begin{equation}
	R_{u} = \frac{P_{u}}{P_{u}(\exper)}\,, \qquad 
	R_{c} = \frac{P_{c}}{P_{c}(\exper)} \label{R-uc}
\end{equation}
and require them to stay below one.

Since we do not assume exact scalar alignment, the SM-like boson $\hsm$
can also possess small FCNCs in the up-quark sector and, consequently, can induce additional top quark decays. 
To take these channels into account, we compute the branching fractions
\begin{equation}
	P_u^{\SM} = \Br(t\rightarrow u \hsm)\,, \qquad									%% previous A5
	P_c^{\SM} = \Br(t\rightarrow c \hsm)\,,											%% previous A6
	\label{br3}
\end{equation}
and compare them with the CMS upper limits from Ref.~\cite{CMS:2021hug} that were based on the $\hsm\to\gamma\gamma$ decay channel: 
$P_u^{\SM}(\exper) < 0.019\%$ and $P_c^{\SM}(\exper) < 0.073\%$.

Finally, for every parameter space point, we combine the above contributions and compute the total top quark decay width:
\begin{eqnarray}
	\Gamma_{t}=\Gamma_{t}^{\SM}+\sum_{i, a}\Gamma(t\rightarrow d_{i}H_{a}^{+})+\sum_{j,B}\Gamma(t\rightarrow u_{j}H_{B})\,.\label{top-width}
\end{eqnarray}
For the SM contribution $\Gamma_{t}^{\SM}$, we adopt $\Gamma_{t}^{\SM}=1.322\ \text{GeV}$ computed in \cite{Gao:2012ja} 
at the next-to-leading order (NLO) in the electroweak corrections and the next-to-next-to-leading order in quantum chromodynamics (QCD) correction.
We neglect the possible interference effects between the SM and non-standard contributions, as they arise only in 
a tiny part of the phase space. Finally, for the experimental result, we adopt the 2024 PDG value
$\Gamma_{t}(\exper)=1.42^{+0.19}_{-0.15}$~GeV listed in \cite{ParticleDataGroup:2024cfk}.

\subsection{Theoretical and experimental constraints}

For the scalar sector scan, we take into account the following constraints. 
\begin{itemize}
	\item 
	The bounded-from-below (BFB) conditions are taken from \cite{Ferreira:2017tvy}. These are sufficient conditions
	and might be too restrictive in some parts of the parameter space. 
	However, since the analytical expressions for necessary and sufficient constraints are not known,
	this seems the best choice available.
	\item
	An explicit algorithm for computing tree-level perturbative unitarity constraints for the CP4 3HDM
	can be found in \cite{Bento:2017eti,Bento:2022vsb}. Since it is computationally expensive, 
	we impose instead a simple prescription to avoid large scalar couplings: 
	we require that the absolute values of all the real parameters $\lambda_{i}$ as well as $\Re\lambda_{8,9}$ and $\Im\lambda_{8,9}$
	do not exceed 5.
	\item
	The electroweak precision test observables $S$, $T$, and $U$ are computed following \cite{Grimus:2008nb}
	and compared with the current PDG values \cite{ParticleDataGroup:2024cfk}:
	\begin{equation}
		S=-0.04\pm0.10 \,,\quad T=0.01\pm0.12 \,,\quad U=-0.01\pm0.09\,,
	\end{equation}
	together with the correlation matrix elements $\rho_{ST} = 0.93$, $\rho_{SU} = -0.70$, $\rho_{TU} = -0.87$.
	In our scan, we select only the points that produce $\chi^2_{STU} < 11.34$, the upper limit corresponding 
	to the 99\% probability with three parameters fitted.
%	\item
%	The properties of the 125~GeV Higgs boson measured by the LHC are taken into account whenever needed.
%	We prefer this manual approach instead of the automated one based on the standard packages such as {\tt HiggsTools}
%	in order to have a better understanding which observables could potentially lead to tensions between the experiment 
%	and the model.
%	The key quantity that we constrain is the angle $\epsilon$ that shows proximity of the model to the scalar alignment
%	regime. Since $\cos\epsilon$ determines the $\hsm ZZ$ and $\hsm WW$ couplings, 
%	we focus on the range $\cos\epsilon > 0.92$.
%	Whenever we monitor $\hsm q\bar q$ couplings, we use the results listed in \cite{ParticleDataGroup:2024cfk}.
\end{itemize}
The CP4 3HDM model is not FCNC-free. Moreover, its Higgs-quark couplings are rather peculiar and do not always follow
the 2HDM intuition. Therefore, it is of paramount importance 
to keep track of the neutral meson oscillation parameters. In the previous work \cite{Zhao:2023hws}, 
this was done according to the simple phenomenological prescription \cite{Nebot:2015wsa}. In that approach, 
constraints were placed on the off-diagonal elements of individual FCNC matrices $N_{d2}$, $N_{d3}$, $N_{u2}$, $N_{u3}$
defined in \eqref{FCNCmatrices}.
However, these matrices are written in the Higgs basis and do not reflect the couplings of the physical Higgs bosons.
Moreover, in this approach, one does not take into account interference between several scalars
contributing to the meson oscillation amplitudes.
In this work, we compute the tree-level contribution to the meson oscillation parameters 
coming from all the neutral scalars with the aid of the expressions from \cite{Babu:2018uik, Babu:2009nn, Babu:2023oih, Deshpande:1993py};
a summary of these formulas is given in Appendix~\ref{appendix-meson}.

\subsection{Scan procedure and parameters: scalar sector} \label{subsection-scan-procedure}

Since the model can display various Higgs spectra, we run the scan in four different regimes.
\begin{itemize}
	\item Sample 1: ``all heavy'', with all new Higgs bosons heavier than the top quark. In this regime,
	the only non-standard top quark decay is induced by the FCNCs of $\hsm$, see Eq.~\eqref{br3}.
	\item Sample 2: ``one charged'', in which we require exactly one charged Higgs boson lighter than the top,
	while all other new Higgs bosons, neutral and charged, stay heavier.
	\item Sample 3: ``one neutral'', in which we demand that, in addition to $\hsm$, there exists exactly one neutral Higgs boson lighter than the top.
	\item Sample 4: ``full sample'', with at least one new Higgs boson lighter than the top quark.
\end{itemize}
Our numerical scan benefits from the two inversion procedures developed in the recent papers
\cite{Liu:2024aew} for the scalar sector and \cite{Zhao:2023hws} for the Yukawa sector.
%These two scans can be decoupled. We first perform a scan in the scalar sector using the twelve parameters \eqref{H-scan}
%and check the compatibility of the generated points with the scalar sector constraints.
%This gives us a sample of $N_{\tiny \rm scalar} = 1000$ models.
We start by sampling the scalar sector within the following ranges of parameters:
\begin{eqnarray}
	&& \tan\beta,\ \tan\psi \in (0.1, \, 10), \quad |\epsilon| < 0.4\,, \quad 
	\alpha, \gamma_1, \gamma_2 \in (0, 2\pi)\,, \quad
	\lambda_{89} \in (0, 5)\,,\nonumber\\
	&& 	%m_{H_1^+} < m_t\,, \quad m_{H_1^+} < 
	50\,\mathrm{GeV} < m_{H_i^+} < 600\,\mathrm{GeV}\,, \quad m_{11}^2 - m_{22}^2 \in (-300^2, 300^2)\,\mathrm{GeV}^2\,.\label{scan-range-scalar}
\end{eqnarray}
The upper limit on $|\epsilon|$ is motivated by the LHC results 
on $\hsm \to VV^*$ ($V$ stands for $W$ or $Z$) \cite{ParticleDataGroup:2024cfk}:
\begin{equation}
\frac{\Gamma^{\LHC}(\hsm\rightarrow WW^*)}{\Gamma^{\SM}(\hsm\rightarrow WW^*)} = 1.00\pm 0.08\,,\quad 
\frac{\Gamma^{\LHC}(\hsm\rightarrow ZZ^*)}{\Gamma^{\SM}(\hsm\rightarrow ZZ^*)} = 1.02\pm 0.08\,.\label{hSM-VV}
\end{equation}
In our model, $\Gamma(\hsm\rightarrow VV^*) / \Gamma^{\SM}(\hsm\rightarrow VV^*) = \cos^2\epsilon$ for both ratios,
so that our upper limit corresponds approximately to this ratio staying inside the $2\sigma$ bound 
from the central LHC values.

However, when scanning within the full range \eqref{scan-range-scalar}, 
we found that in almost all cases the Higgs spectrum features one or several Higgs bosons
lighter than the top quark. This is consistent with the previous scans \cite{Ferreira:2017tvy,Liu:2024aew},
but it makes the search of ``all heavy'' and ``one neutral'' points extremely inefficient.
Fortunately, as was found in \cite{Liu:2024aew},
the probability of getting ``all heavy'' points dramatically increases if we focus on a restricted parameter space,
in which the angles are constrained to lie in the following ranges:
\begin{equation}
	\tan\beta \in [0.5, 2]\,, \quad \tan\psi \in [0.5, 3]\,, \quad
	|\tan\alpha| < 0.05\,,
	\quad |\tan\gamma_{1}|, \ |\tan\gamma_{2}| < 0.3\,.\quad
	\label{focused-scan-range}
\end{equation}
Such a restricted scan generated abundant samples in all four regimes, and we adopted it as our main scan strategy
for all four samples.

\subsection{Scan procedure and parameters: Yukawa sector}

We scan the quark Yukawa sector following \cite{Zhao:2023hws}. 
This procedure starts with the physical quark masses and mixing parameters as input, 
together with the vevs generated at the previous step, and reconstructs
the Yukawa matrices that correspond to the scenarios $(A,B_2)$ or $(B_1, B_1)$, 
the two options deemed viable in \cite{Zhao:2023hws}.
By construction, the Yukawa sector generated in this way reproduces all the quark masses and mixing angles, 
including the $CP$-violating phase. 

Which quark masses should be used as input is a subtle point that requires discussion.
When one fits quark properties, one usually aims to reproduce pole masses for heavy quarks 
and low-scale masses for light quarks, such as the values listed by the PDG \cite{ParticleDataGroup:2024cfk}:
\begin{align}
	\label{masses-pole}
	\begin{split}
		&m_u(2\ \text{GeV})=2.16\ \text{MeV}\,,\quad m_c(m_c)=1.273\ \text{GeV}\,,\quad m_t(m_t)= 172.4\ \text{GeV}\,,\\
		&m_d(2 \ \text{GeV})=4.7\ \text{MeV}\,,\quad m_s(2 \ \text{GeV})=93.5\ \text{MeV}\,,\quad m_b(m_b)=4.183\  \text{GeV}\,.
%		&m_W=80.3692\ \text{GeV}\,,\quad m_Z=91.1880\ \text{GeV}
	\end{split}
\end{align}
However, the quark masses exhibit substantial renormalization group (RG) evolution, and their values
at a high energy scale such as $m_Z$ are reduced \cite{Xing:2007fb}:
\begin{align}
	\label{masses-mZ}
	\begin{split}
		&m_u(m_Z)=1.27\ \text{MeV}\,,\quad m_c(m_Z)=0.619\ \text{GeV}\,,\quad m_t(m_Z)=171.7 \ \text{GeV}\,,\\
		&m_d(m_Z)=2.90\ \text{MeV}\,,\quad m_s(m_Z)=0.055\ \text{GeV}\,,\quad m_b(m_Z)=2.89\  \text{GeV}\,.
	\end{split}
\end{align}
Quark masses come from Yukawa couplings, which implies that these couplings
are significantly reduced at a high energy scale.
In particular, when one computes within the Standard Model the Higgs boson decays to quark pairs, 
the relevant Yukawa coupling should be taken at a high energy scale; 
otherwise, one would face uncomfortably large negative QCD loop corrections \cite{Djouadi:2005gi}.

Now comes a subtle point. In the SM or in simple multi-Higgs models such as Type-II 2HDM,
the physical Yukawa couplings are expressible in terms of quark masses.
As a result, when one computes Higgs to quark decays, one can just use the RG-evoled quark masses 
in the matrix elements. However, in our model, as well as in a generic multi-Higgs-doublet model
without natural flavor conservation, one has many relevant Yukawa couplings, which are not expressible 
in terms of very few quark masses. What values should one take when calculating the numerous
top quark or Higgs boson decays listed above?

A proper treatment would be to compute the RG evolution of the entire Yukawa sector 
for each parameter space point and to use in these decays the evolved values of the relevant couplings.
This procedure is computationally expensive and is not particularly rewarding.
We adopted a different method, which we believe offers a good approximation to this procedure.
Namely, when we generate a CP4 3HDM Yukawa sector, we use, from the very beginning, 
not the pole quark masses \eqref{masses-pole} but the high-scale quark masses \eqref{masses-mZ}.
In this way, all coupling matrices of the resulting Yukawa sector, both the diagonal $M_d$, $M_u$ 
and the FCNC matrices $N$'s, are taken by construction at the high scale 
and can be directly inserted into the decay amplitudes without further RG evolution.
%One should be careful not to overdo this procedure when RG evolution is not involved.
%In particular, when computing the tree-level Higgs boson contributions to the meson oscillation parameters,
%see Appendix~\ref{appendix-meson}, we use quark pole masses, not their high-scale values.
Finally, as for the CKM matrix parameters, 
RG evolution is slow, and we just use the standard PDG values \cite{ParticleDataGroup:2024cfk}:
\begin{equation}
	\sin\theta^{\CKM}_{12}=0.22501\,,\quad \sin\theta^{\CKM}_{23}=0.04183\,,\quad \sin\theta^{\CKM}_{13}=0.003732\,,\quad
	\delta = 1.147\,.
\end{equation}

When performing the Yukawa sector scan, we use as input not only the physical quark masses and mixing parameters
but also the angles of the right-handed quark rotation matrices, see details in \cite{Zhao:2023hws}.
In particular, the angles $\theta_{12}$ and $\theta_{13}$ in all relevant matrices,
as well as the angle $\theta_{23}$ for the case $(B_1,B_1)$, are sampled uniformly between $-\theta_{max}$ and $\theta_{max}$, where
$\theta_{max}$ can be set equal to $\pi$ (a full Yukawa scan) 
or a small parameter such as $\pi/10^4$ or lower (a restricted Yukawa scan).

We resort to the restricted scan out of necessity to bring the neutral meson oscillation parameters under constraints.
As was already demonstrated in \cite{Zhao:2023hws}, generic rotation angles
lead to large FCNC couplings, which generate prohibitively strong neutral meson oscillations,
while a restricted Yukawa scan can suppress these contributions.
However, the work \cite{Zhao:2023hws} did not use the full scalar sector and  
relied on a simple procedure borrowed from \cite{Nebot:2015wsa} to control this influence.
Namely, the off-diagonal elements of the individual matrices $N_{d2,d3}$ and $N_{u2,u3}$ were directly computed 
and upper limits were placed on their values assuming a heavy Higgs boson with a mass of 1~TeV.
However, as found in \cite{Liu:2024aew}, CP4 3HDM does not accommodate such heavy scalars. 
This is why this simplistic approach is insufficient and a proper calculation of non-standard contributions 
to the meson oscillation parameters are needed.
This is what we now perform; details of this computation are given in Appendix~\ref{appendix-meson}.

\begin{figure}[h!]
	\centering
	\includegraphics[width=0.8\textwidth]{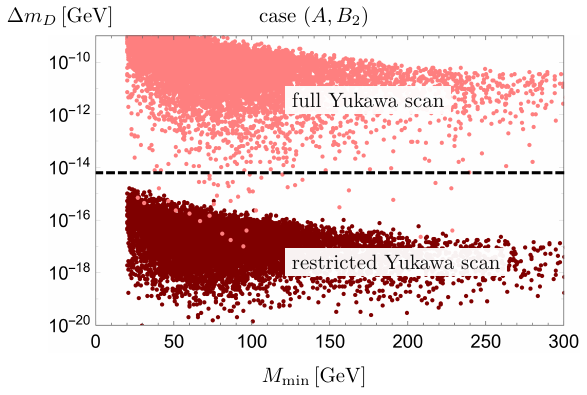}
	\caption{The contribution to the neutral $D$-meson mass splitting $\Delta m_D$ coming from the tree-level FCNC
		as a function of $M_{\min}$, the minimal mass of any non-SM Higgs boson. 
		The Yukawa sector is of type $(A,B_2)$. The light pink points correspond to the full Yukawa scan,
		while the dark red points emerge from the restricted Yukawa scan with $\theta_{max} = \pi/10^4$.
		The dashred line corresponds to the experimentally measured $\Delta m_D$.}
	\label{fig-D-meson}
\end{figure}

For the Yukawa sector $(A,B_2)$, we only need to take care of $D$-meson oscillations.
We found that in a general Yukawa scan the tree-level FCNC contribution 
to the neutral $D$-meson mass splitting  $\Delta m_{D}$ 
is typically large, often orders of magnitudes larger than the experimentally measured value 
$\Delta m_{D} = 6.25\times10^{-15}$ GeV.
This is shown in Fig.~\ref{fig-D-meson} by light pink points.
However, restricting $\theta_{12}$ and $\theta_{13}$ in $V_{uR}$ by $\theta_{max} = \pi/10^4$ strongly 
suppresses this contribution even for light $M_{\min}$, the minimal mass of any non-SM Higgs boson.
These non-standard contributions are shown in Fig.~\ref{fig-D-meson} by dark red points.
They lie significantly below the experimental value, which means that their effect can be safely neglected.
We have also verified that a simplistic check in the spirit of \cite{Zhao:2023hws} 
is in agreement with these results.

The bottom line is that we can now proceed with the top decays within the scenario $(A,B_2)$, 
and as long as we use the restricted Yukawa scan with $\theta_{max} = \pi/10^4$, 
we no longer need to worry about $D$-meson oscillations. 

\subsection{Yukawa sector $(B_1,B_1)$ is ruled out}

\begin{figure}[h!]
	\centering
	\includegraphics[width=0.8\textwidth]{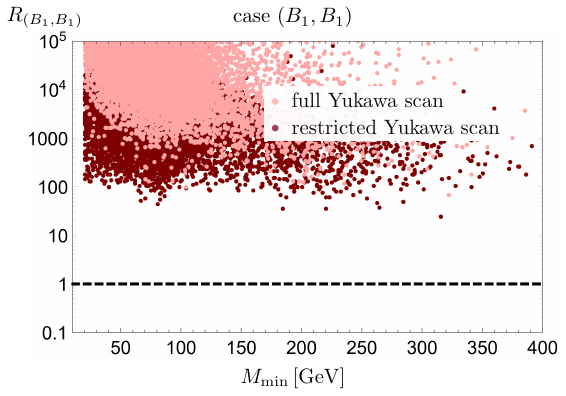}
	\caption{The quantity $R_{(B_1,B_1)}$ defined in Eq.~\eqref{RB1B1} that shows the 
		magnitude of the tree-level FCNC contributions to the neutral meson oscillation parameters
		relative to their experimental values.
		The Yukawa sector is of type $(B_1,B_1)$; the labels are the same as in Fig.~\ref{fig-D-meson}.
		The dashed line corresponds to $R_{(B_1,B_1)} = 1$, the target value for this ratio.
		}
	\label{fig-B1B1-dead}
\end{figure}

When repeating the same meson oscillation check for the Yukawa scenario $(B_1,B_1)$,
we discovered that the conclusion of \cite{Zhao:2023hws} concerning this case was too optimistic.
In this Yukawa scenario, the tree-level FCNCs take place in both up and down sectors.
Thus, one must check whether the non-standard contributions do not exceed experimental limits
for $K$, $B$, $B_s$, and $D$-meson oscillations.
For a point in the parameter space, we compute all the tree-level contributions to the neutral meson oscillations,
divide them by the corresponding experimental values, and require that all these ratios be less than 1.
We combine all these checks by defining $R_{(B_1,B_1)}$ as the largest among these quantities:
\begin{eqnarray}
	R_{(B_1,B_1)}=\max\left[\frac{\Delta m_{D}}{(\Delta m_{D})_{\exper}},
	\frac{\Delta m_{K}}{(\Delta m_{K})_{\exper}}, 
	\frac{|\varepsilon_{K}|}{(|\varepsilon_{K}|)_{\exper}}, 
	\frac{\Delta m_{B_{s}}}{(\Delta m_{B_{s}})_{\exper}}, 
	\frac{\Delta m_{B_{d}}}{(\Delta m_{B_{d}})_{\exper}}\right]\,. \label{RB1B1}
\end{eqnarray}
This quantity is plotted in Fig.~\ref{fig-B1B1-dead} as a function of the same $M_{\min}$, 
the mass of the lightest non-SM Higgs boson. 
As before, the light pink points correspond to the full Yukawa scan, 
while the dark red points are for $\theta_{max}$ set as low as $\pi/10^7$.
We see that all points lead to $R_{(B_1,B_1)} > 1$, often exceeding it by orders of magnitude, 
in the entire mass range of $M_{\min}$. This distribution does not improve
within the restricted Yukawa scan. The explanation is simple: in the down-quark sector,
kaon, $B$, and $B_s$-meson oscillations probe all the off-diagonal entries of the matrices $N_{d2}$, $N_{d3}$, 
and there is no limit in which these FCNC elements would all be suppressed.

Arguably, this procedure is not as strict as it should be.
Indeed, since these are not the upper limits but the measured values,
they receive both the SM and non-standard contributions. 
Thus, the room for any non-standard contribution should be even smaller.
However, even with such a generous margin, we cannot find any suitable $(B_1,B_1)$ point  
that would satisfy all the meson oscillation constraints.
The conclusion is that neutral meson oscillations definitely rule out 
the CP4-symmetric Yukawa scenario $(B_1,B_1)$.
In the rest of the paper, we will only deal with the sole surviving case, scenario $(A, B_2)$.

\section{Yukawa sector $(A,B_2)$: numerical results} 
\label{section-scan-results}

\subsection{Constraints coming from $\hsm t\bar t$ interactions}

As described in section~\ref{subsection-scan-procedure},
we scan the parameter space under four different regimes
depending on how many new Higgs bosons are lighter than the top quark.
However, in all these cases, we have the SM-like Higgs particle $\hsm$
that can couple to the top quark in non-standard ways, and we begin by inspecting its signals.

In the Yukawa sector $(A,B_2)$, $\hsm$ couples to down quarks in the SM-like way,
but the up-quarks are governed by 
$\overline{u}_{L}\widetilde{N}_{u\SM}u_{R}\,\hsm/v+H.c.$
The coupling matrix $\widetilde{N}_{u\SM}$ comes from the second line of Eq.~\eqref{L-n-c} together 
with Eq.~\eqref{L-n-p2}:
\begin{equation}
	\widetilde{N}_{u\SM}=M_{u} c_\epsilon 
	+ s_\epsilon \left[N_{u2}(c_\alpha c_{\gamma_1}-i s_\alpha s_{\gamma_2}) 
	+ N_{u3}(c_\alpha s_{\gamma_1}-i s_\alpha c_{\gamma_2})\right]\,.
	\label{L-n-c-again}
\end{equation}
%Here, $M_u = \diag(m_u, m_c, m_t)$, since we work already in the basis of physical quarks.
The generic structure of the Yukawa matrices $N_{u2}$ and $N_{u3}$ was described in \cite{Zhao:2023hws}.
For the case $B_2$ sampled in a restricted Yukawa scan, it reveals the following pattern: 
the largest element in $N_{u2}$ is the diagonal element $N_{u2, tt}$, of the order of $m_t$, 
while the largest element in $N_{u3}$ is the off-diagonal elements $N_{u3, tu}$,
of the same order.
These two large elements, $N_{u2, tt}$ and $N_{u3, tu}$, appear in $\widetilde{N}_{u\SM}$
with the suppressing factor $\sin\epsilon$ times the sines or cosines of other angles.
For a generic situation far from the alignment limit, with $\epsilon \sim 1$,
one expects a large modification of the $\bar t t \hsm$ coupling relative to its SM value, 
usually denoted by $\kappa_t$,
as well as a very sizable non-standard $\bar t u \hsm$ coupling, 
which leads to the decay $t \to \hsm u$, whose branching fraction is denoted as $P_u^{\SM}$.

\begin{figure}[h!]
	\centering
	\includegraphics[width=0.8\textwidth]{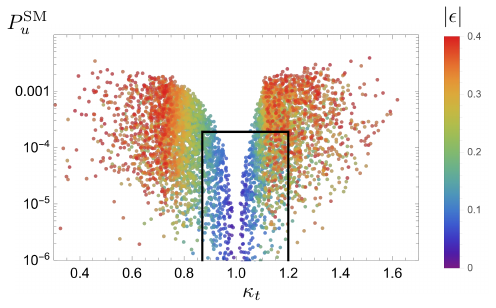}
	\caption{The Higgs-top coupling relative to its SM value ($\kappa_t$) and the branching fraction of $t\to \hsm u$
		in a restricted scalar and Yukawa scans. The color encodes the magnitude of $|\epsilon|$, 
		with larger-$|\epsilon|$ points plotted on top of lower-$|\epsilon|$ ones.  
		The box corresponds to the tightest LHC bounds.}
	\label{fig-hsm}
\end{figure}

In Fig.~\ref{fig-hsm}, we show the distribution of these two observables in a restricted scalar and restricted Yukawa scan,
with color showing the value of $|\epsilon| \le 0.4$.
The box corresponds to the experimental values: 
$\kappa_t \in (0.87, 1.20)$ as reported in \cite{ATLAS:2022tnm},
while the upper limit on $P_u^{\SM}$ is $1.9\times 10^{-4}$ taken from \cite{CMS:2021hug}.
As expected, for very small $|\epsilon| < 0.04$, 
all points lie within the experimental limits for any choice of the other angles.
As $|\epsilon|$ grows, an increasingly larger fraction of the points fall outside the box,
and for $|\epsilon| > 0.2$ ($\cos\epsilon < 0.98$),
most points violate one or both limits.
This implies that $\kappa_t$ and $t \to \hsm u$ place stronger constraints on the CP4 3HDM parameter space
than the decays $\hsm \to VV$, see Eq.~\eqref{hSM-VV}, and must be taken into account in our subsequent analysis.

\subsection{The impact of the total top decay width}

\begin{figure}[h!]
	\centering
	\includegraphics[width=0.48\textwidth]{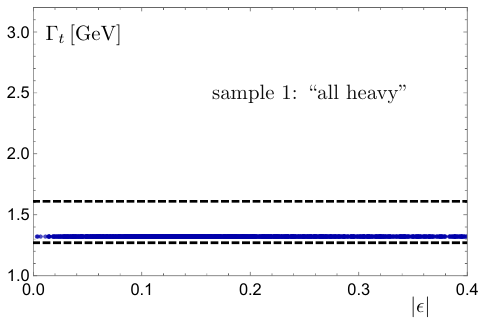}
	\includegraphics[width=0.48\textwidth]{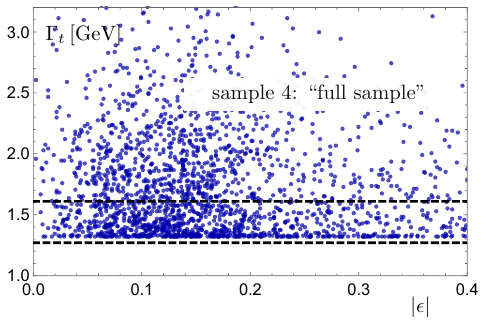} \\
	\includegraphics[width=0.48\textwidth]{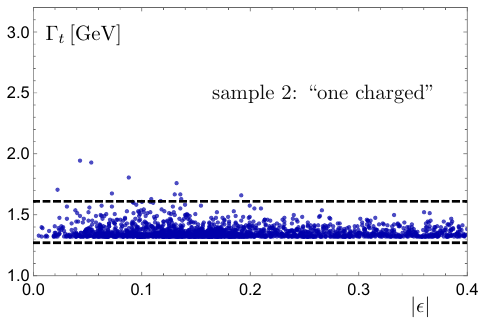}
	\includegraphics[width=0.48\textwidth]{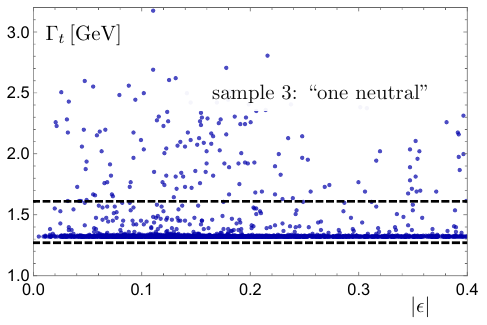} 
	\caption{The total top quark decay width as given in Eq.~\eqref{top-width} for the case $(A, B_{2})$. 
		The four panels correspond to the four scalar sector samples listed in section~\ref{subsection-scan-procedure}.
		The band between the dashed lines corresponds to the $1\sigma$ experimental range established by PDG \cite{ParticleDataGroup:2024cfk}.}
	\label{fig-total-width}
\end{figure}

Next, we turn to the total top quark decay width $\Gamma_t$,
which receives non-standard contributions from decays to all light Higgs bosons,
including the SM-like Higgs particle.

In Fig.~\ref{fig-total-width}, we plot $\Gamma_t$ against $|\epsilon|$. 
At this point, we already impose the constraints on $\kappa_t$ described above.
The four panes correspond to the four regimes listed in section~\ref{subsection-scan-procedure}.
In the ``all heavy'' sample, the top left panel of Fig.~\ref{fig-total-width}, 
only the SM-like Higgs gives non-SM contributions to $\Gamma_t$ 
through the decays $t\rightarrow u \hsm$ and $t\rightarrow c \hsm$. As expected, their effect on $\Gamma_t$ is negligible.
We see a completely different situation in the ``full sample'', the top right panel of Fig.~\ref{fig-total-width},
where the new Higgs bosons significantly modify the total top quark width. 
The effect persists for all values of $\epsilon$ and survives in the alignment limit, 
which was reported already in \cite{Ivanov:2021pnr}.
We checked that changing the details of the Yukawa parameter scan (full or restricted) does not alter this distribution.
As illustrated by the lower row of the plots, 
the largest contributions typically come from neutral light Higgs bosons.

Thus, we keep only the points passing the $\Gamma_t$ constraint and proceed further. 

\subsection{The impact of extra light Higgs bosons on top decays}

We now turn to the signatures induced by the top quark decays to new scalars. 
Let us first consider the scenario with only one charged Higgs boson, which we denote simply as $H^+$.
For each point of the scan, we calculate the decay width $t\rightarrow bH^{+}$
and all available quark pair decays $H^{+}\rightarrow u_i \overline{d}_j$. 
We then compute the branching fractions $\Br(H^{+}\rightarrow c\overline{b})$ and 
$\Br(H^{+}\rightarrow c\overline{s})$, find the branching ratio products
$P_{cb}$ and $P_{cs}$ given in Eq.~\eqref{br1}, and finally calculate their ratios
to the experimental upper limits found by the LHC for the mass of $H^+$, 
$R_{cb}$ and $R_{cs}$ defined in Eq.~\eqref{R-cbcs}.
Only the points that satisfy $R_{cb} < 1$ and $R_{cs} < 1$ pass this check. 

\begin{figure}[h!]
	\centering
	\includegraphics[width=0.49\textwidth]{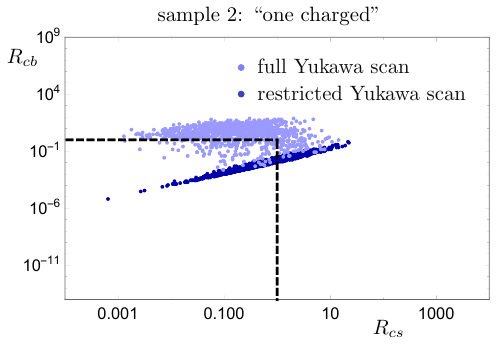}
	\includegraphics[width=0.49\textwidth]{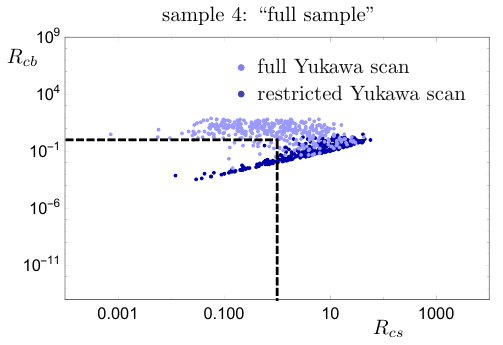}
	\caption{The impact of light charged Higgs bosons on the non-standard top quark decays.
		Shown are the ratios of the decay chain probabilities, ending in $H^+ \to c\bar b$ and $H^+ \bar s$, 
		to the corresponding experimental upper limits defined in \eqref{R-cbcs}.
		Light and dark blue points correspond to the results of a full and restricted Yukawa scan, respectively.
		The left panel contains only one charged Higgs boson, the right panel corresponds to the full sample.
		Only points with both $R_{cb} < 1$ and $R_{cs} < 1$, marked by the dashed lines, pass this check. 
	}
	\label{fig-R-cbcs}
\end{figure}

In Fig.~\ref{fig-R-cbcs}, left, we present this scatter plot for different scan options. 
All points shown here pass the $\hsm$ and $\Gamma_t$ constraints discussed in the previous sections.
The light blue points emerge from a full Yukawa scan with unconstrained angles;
the dark blue points come from a restricted scan with $\theta_{\max} = \pi/10^4$.
We see that restricting Yukawa rotation angles not only keeps the $D$-meson oscillations
standard but also helps bring $\Br(H^{+}\rightarrow c\overline{b})$ under control.

In Fig.~\ref{fig-R-cbcs}, right, we repeat this study for the ``full sample.''
In this case, whenever we have two light charged Higgs bosons, we compute $R_{cb}$ and $R_{cs}$ for both of them
and put a point corresponding to the largest values of $R_{cb}$ and $R_{cs}$.
As expected, the FCNC decays are now stronger, with fewer points that pass both constraints. 

\begin{figure}[h!]
	\centering
	\includegraphics[width=0.49\textwidth]{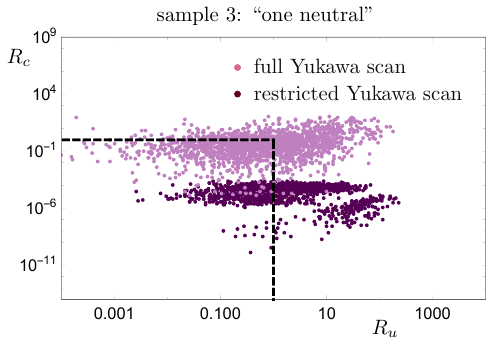}
	\includegraphics[width=0.49\textwidth]{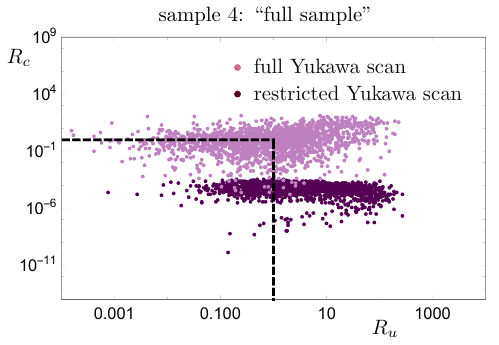}
	\caption{The impact of light neutral Higgs bosons on non-standard top quark decays 
		in the ``one neutral'' (left) and ``full sample'' (right) regimes.
		Shown are the distributions of the ratios $R_{u}$ and $R_{c}$ defined in Eq.~\eqref{R-uc}
		as branching ratio predictions relative to the experimental upper bounds.
		Only points with both $R_{u} < 1$ and $R_{c} < 1$, showed by the dashed lines, pass this check. 
	}
	\label{fig-R-cu}
\end{figure}

Light neutral Higgs bosons can also produce a strong signal in top FCNC decays. 
We now perform a scan in the ``one neutral'' regime and denote the light neutral scalar by $H$.
For each point, we compute $\Br(t\rightarrow uH)$ and $\Br(t\rightarrow cH)$,
as well as $\Br(H\rightarrow b\overline{b})$. Then we construct the ratios
$P_u$ and $P_c$ of our predictions to the experimental upper limits, see Eq.~\eqref{R-uc}. 
Both $R_{u} < 1$ and $R_{c} < 1$ are required for a point to pass this check. 

In Fig.~\ref{fig-R-cu}, left, we show the result of this comparison.
Again, all points pass the $\hsm$ and $\Gamma_t$ constraints.
The light violet points emerge from a full Yukawa scan, while
the dark violet points come from a restricted scan with $\theta_{\max} = \pi/10^4$.
In the restricted Yukawa scan, the $t\overline{c}H$ coupling is suppressed,
but the $t\overline{u}H$ remains rather large, so that only a part of the points survive.
This is compatible with our general observation:
in the CP4 3HDM, there is no limit in which all FCNC couplings are suppressed;
even if $\hsm$ is FCNC-free, the other Higgs bosons induce a sizable $t\to u$
transition. 
Fortunately, this effect is not extreme, and a part of the scanned 
parameter space stays within experimental limits.
The ``full sample,'' illustrated by Fig.~\ref{fig-R-cu}, right, 
shows a similar distribution.

\subsection{Overall results and benchmark points}

Having discussed the impact of various constraints individually,
we now look for viable CP4 3HDM models within the scenario $(A,B_2)$ that pass all the checks:
the theoretical and electroweak precision constraints on the scalar sector, 
the contribution to $\Delta m_D$ not exceeding the measured value,
the constraints from $\hsm$ and $\Gamma_t$, 
and the top quark decays to lighter scalars, which must stay sufficiently suppressed.

Guided by the above insights, we found viable points for all four scan regimes.
For the ``all heavy'' case, this was rather straightforward:
staying sufficiently close to the alignment limit, 
we could easily keep all the points inside the  dashed box in Fig.~\ref{fig-hsm}.
But even in the presence of additional charged or neutral Higgs bosons,
we generated a sizable sample of points passing all the constraints studied here.
In Appendix~\ref{appendix-benchmark}, we present two benchmark points with the following properties.
\begin{itemize}
	\item 
	Benchmark point 1, scenario ``one charged,'' corresponds to $\epsilon = - 0.207$ ($\cos\epsilon =0.9787$).
	The charged Higgs boson masses are $149.1$ and $248.4$~GeV, the neutral Higgs boson masses are
	\begin{equation}
		m_{H_A}\,[\mbox{GeV}] = 125.0,\ 186.4,\ 209.6,\ 235.5,\ 324.8.
	\end{equation}
	The $D$-meson mass splitting is $\Delta m_D = 8.02\times 10^{-18}$~GeV.
	The top to SM Higgs boson decays have the probabilities $\Br(t \to \hsm u) = 5.1\times 10^{-5}$
	and $\Br(t \to \hsm c) = 8.4\times 10^{-10}$.
	The $\hsm t\bar t$ coupling modifier is $\kappa_t = 1.089$, the total top width is $\Gamma_{t}=1.351$~GeV.
	The signals due to top to charged Higgs decays are $P_{cb} = 8.64 \times 10^{-6}$
	and $P_{cs} = 2.17\times 10^{-4}$. 
	\item 
	Benchmark point 2, scenario ``one neutral'', corresponds to $\epsilon = 0.124$ ($\cos\epsilon = 0.9923$).
	The charged Higgs boson masses are $258.9$ and $305.0$ GeV, the neutral Higgs boson masses are
	\begin{equation}
		m_{H_A}\,[\mbox{GeV}] = 125.0,\ 145.7,\ 190.2,\ 265.3,\ 268.7.
	\end{equation}
	The $D$-meson mass splitting is $\Delta m_D = 9.80\times 10^{-18}$~GeV.
	The top to SM Higgs boson decays have the probabilities $\Br(t \to \hsm u) = 1.3\times 10^{-4}$
	and $\Br(t \to \hsm c) = 3.3\times10^{-10}$.
	The $\hsm t\bar t$ coupling modifier is $\kappa_t = 0.922$, the total top width is $\Gamma_{t}=1.409$~GeV.
	The signals due to top to neutral Higgs decays are $P_{c} = 4.6\times10^{-12}$
	and $P_{u} = 5.4\times 10^{-5}$.
\end{itemize}
For both benchmark points, all of the observables checked stay within the corresponding experimental limits.
%We find this a non-trivial result, taking into account the tight correlations among the scalar and Yukawa
%sector and the unavoidable presence of FCNCs within the CP4 3HDM.
%We welcome the community to explore other phenomenological 

\section{Discussion and conclusions}\label{section-conclusion}

CP4 3HDM is a unique multi-Higgs model that implements an exact $CP$-symmetry of order 4 without any additional assumption
\cite{Ivanov:2015mwl}.
This single symmetry shapes the scalar and Yukawa sectors in a very particular way, limiting the number of free parameters
and inducing non-trivial correlations among the scalar and Yukawa sectors.
Previous studies \cite{Ferreira:2017tvy,Zhao:2023hws,Liu:2024aew} found that there is a part in the parameter
space that leads to a reasonable scalar sector, can reproduce all the quark masses and mixing patterns, 
including the correct amount of $CP$ violation, and can even keep the tree-level contributions 
to the neutral meson oscillation parameters under control.
It was a non-trivial task to satisfy all these constraints, especially taking into account that
there is no decoupling limit of the model and that it is impossible to completely avoid FCNC in all sectors.
Moreover, the FCNC patterns emerging in this model are rather peculiar and do not follow the 2HDM-based intuition.

Thus, the phenomenological fate of this model remains unclear. Can it survive, at least in a corner of the parameter space,
all the present-day constraints from the scalar and flavor experiments, or is it eventually ruled out by some of them?

In this work, we took a further step in this analysis. We investigated how top quarks
couple to all the Higgs bosons and what constraints on the model parameter space follow.
We studied several regimes: ``all heavy'' (all bosons apart from $\hsm$ are heavier than the top quark),
``one charged'' and ``one neutral'', in which we required exactly one charged or neutral scalar to be ligher than the top,
and the ``full sample'' regime, in which we simply assume that there is at least one new boson lighter than the top quark.
The observables taken into account were the total top quark width $\Gamma_t$, non-standard top-quark decays
to the SM-like Higgs $t \to u\hsm$ and $t \to c\hsm$, the magnitude of the $\hsm t\bar t$ coupling,
and the various non-standard decays of the top quark to lighter neutral and charged Higgs bosons that were searched for
at the LHC.

By studying the impact of these observables and performing an overall scan in the scalar and Yukawa parameter spaces, 
we made the following observations.
\begin{itemize}
	\item 
	To keep the $\hsm$-top couplings under control, we need to stay close to the alignment limit (the angle $\epsilon \ll 1$).
	These observables constrain $\epsilon$ more strongly than the decays $\hsm \to WW^*$ and $\hsm \to ZZ^*$.
	\item
	When performing a scan, it is highly beneficial to employ the so-called inversion procedure devised for the Yukawa
	sector in \cite{Zhao:2023hws} and for the scalar sector in \cite{Liu:2024aew}. This procedure uses several observables as input
	and reconstructs the parameters of the lagrangian. 
	The standard scanning procedure (picking a random seed point for lagrangian parameters 
	and trying to optimize the global $\chi^2$) is extremely inefficient and does not
	yield even a single viable point in reasonable time.
	\item
	We found that the CP4 invariant Yukawa scenario $(B_1,B_1)$, which was identified in the previous work \cite{Zhao:2023hws}
	as potentially viable, is ruled out by the combination of all neutral meson oscillation parameters.
\end{itemize}
The only Yukawa scenario that can satisfy all the constraints is $(A,B_2)$.
In this case, there are no tree-level FCNC effects in the down-quark sector,
while the contribution to the $D$-meson mass splitting can be parametrically suppressed.
At the same time, non-standard couplings of the $\hsm$ to top quarks can be also easily controlled
by the angle $\epsilon$. Thus, it was easy to generate a large sample of points in the ``all heavy'' regime,
in which all extra Higgs bosons are heavier than the top quark.

When one or more light Higgs bosons are present, FCNC top quark decays
cannot be eliminated. However, there remain parts of the parameter space
that safely pass all the constraints we considered.
We generated samples of viable points in all the regimes considered.
For illustration, we presented two benchmark points, 
one with a light charged Higgs particle and another with a light additional neutral boson.
For these benchmark points, all the observables studied in this work are within experimental limits.

In summary, using several top quark related observables, 
we managed to further corner the viable parameter space of the CP4 3HDM, 
but we still could not rule it out.
As we now know how to focus on the most promising part of the parameter space,
we can feed the output of our scan to {\tt HiggsTools} or other packages 
and verify whether any points survive. This will be the objective of a future work.

\section*{Acknowledgments}

This work was supported by Guangdong Natural Science Foundation (project No. 2024A1515012789).

\appendix

\section{Non-$q\bar q$ decays of neutral scalars}\label{appendix-nonqq}

In section~\ref{subsection-top-decays}, we listed various Higgs boson decays to quark pairs.
Each bosons can, of course, decay to other final states.
For light charged Higgs bosons, we neglect other channels. If it happens that additional channels
such as $H^+_a \to \tau^+\nu$ are important, the FCNC signals discussed in this work will be further suppressed.
Whether these non-$q\bar q$ channels could constitute another strong handle on the CP4 3HDM requires 
a dedicated study of the lepton Yukawa sector.

For light neutral Higgs bosons $H_A$, there are decays to gauge boson pairs that are not 
{\em \`a priori} small and can enhance the total decay width of each scalar. 
As a result, the $q_i\bar q_j$ branching ratios and their observable signals can be suppressed, 
possibly re-opening some part of the parameter space.
Therefore, when computing the $q_i\bar q_j$ branching ratios of neutral scalars, 
we also evaluated the following decays: $\Gamma(H_A\rightarrow VV)$, where $VV = WW^{\star}$ and $ZZ^{\star}$, 
and $\Gamma(H_A\rightarrow gg)$.

For $\Gamma(H_A\rightarrow VV)$, we relate it to the SM Higgs decay width taken at the mass $m_{H_A}$ instead of 125 GeV via 
\begin{equation}
	\Gamma(H_A\rightarrow VV)=S_{1A}^{2}\cdot\Gamma(\hsm(m_{H_A})\rightarrow VV)\,.
\end{equation}
The values of the SM Higgs decays at different masses were taken from \cite{LHCHiggsCrossSectionWorkingGroup:2013rie}.
Since we work close to the scalar alignment, this decay width is suppressed 
by the coupling $S_{1A}^{2} < \sin^2\epsilon \ll 1$, but the SM decay width itself shoots up
in the mass range 140--160 GeV. Indeed, we observed that sometimes $\Gamma(H_A\rightarrow VV)$ becomes
the dominant decay channel near the upper end of this range.

For $\Gamma(H_{A}\rightarrow gg)$, we relate it with the SM expression \cite{Djouadi:2005gi}:
\begin{equation}
\Gamma(H_A\rightarrow gg) = K_{\mathrm{NLO}}\cdot \frac{G_{F}\alpha_{s}^{2}m_{H_A}^{3}}{36\sqrt{2}\pi^{3}}
\left|\frac{3}{4}\left(\frac{\widetilde{N}_{dA,bb}}{m_{b}}A_{1/2}(\tau_{b})
+\frac{\widetilde{N}_{uA,tt}}{m_{t}}A_{1/2}(\tau_{t})\right)\right|^{2}\,.
\end{equation}
Here, we adopt the fixed-order prescription outlined in \cite{Djouadi:2005gi}: 
inside the loop functions, we use the pole masses for the $b$ and $t$ quarks but explicitly include the $K$-factor that
takes care of the NLO QCD corrections: $K_{\mathrm{NLO}} = 1 + 215\alpha_s(m_{H_A})/(12\pi)$.
The loop function $A_{1/2}(\tau_{q})$ has the form
\begin{equation}
A_{1/2}(\tau_{q})=\frac{2[\tau_{q}+(\tau_{q}-1)f(\tau_{q})]}{\tau_{q}^{2}}\,,
\end{equation}
with $\tau_{q}=m_{H_A}^{2}/(4m_{q}^{2})$ and 
\begin{equation}
f(\tau_{q}) =
\arcsin^{2}\!\sqrt{\tau_{q}} \quad \mbox{for}\quad \tau_{q} \leq 1\,,\quad 
-\frac{1}{4}\left(\log\frac{1+\sqrt{1-\tau_{q}^{-1}}}{1-\sqrt{1-\tau_{q}^{-1}}}-i\pi\right)^{2} 
\quad \mbox{for}\quad \tau_{q} > 1\,.
\end{equation}

\section{Tree-level contributions to the neutral meson mixing}\label{appendix-meson}

We compute the tree-level contributions of the neutral Higgs bosons to meson oscillation parameters
using the prescriptions described in \cite{Babu:2018uik, Babu:2009nn, Babu:2023oih, Deshpande:1993py}.
The effective Hamiltonian for $\Delta F=2$ processes has the form
\begin{align}
	\mathcal{H}_{\mathrm{eff}}=-\sum_A\frac{1}{2 m_{A}^{2}}\left(\bar{q}_{i}\left[\mathbb{Y}_{q,ij}^{A} \frac{1+\gamma_{5}}{2}
	+ (\mathbb{Y}_{q}^{A\dagger})_{ij} \frac{1-\gamma_{5}}{2}\right] q_{j}\right)^{2}.
\end{align}
where $m_{A}$ is the mass of each neutral scalar $H_A$ and $\mathbb{Y}^A_{q,ij}=\sqrt{2}\widetilde{N}_{qA, ij}/v$.
In the modified vacuum saturation and factorization approximation,
\begin{align}
	\left\langle\phi\left|\bar{f}_{i}\left(1 \pm \gamma_{5}\right) f_{j} \bar{f}_{i}\left(1 \mp \gamma_{5}\right) f_{j}\right| \bar{\phi}\right\rangle&=f_{\phi}^{2} m_{\phi}\left(\frac{1}{6}+\frac{m_{\phi}^{2}}{\left(m_{i}+m_{j}\right)^{2}}\right) \cdot B_{4}, \\
	\left\langle\phi\left|\bar{f}_{i}\left(1 \pm \gamma_{5}\right) f_{j} \bar{f}_{i}\left(1 \pm \gamma_{5}\right) f_{j}\right| \bar{\phi}\right\rangle&=-\frac{5}{6} f_{\phi}^{2} m_{\phi} \frac{m_{\phi}^{2}}{\left(m_{i}+m_{j}\right)^{2}} \cdot B_{2},
\end{align}
the relevant matrix element for the meson transition $\bar\phi \to \phi$ is 
\begin{align}
	\mathcal{M}_{12}^{\phi}=\left\langle\phi\left|H_{\mathrm{eff}}\right|\bar{\phi}\right\rangle
	= -\sum_A\frac{f_{\phi}^{2} m_{\phi}}{2 m_{A}^{2}}
	&\left[-\frac{5}{24} \frac{m_{\phi}^{2}}{\left(m_{i}+m_{j}\right)^{2}}
	\left((\mathbb{Y}_{q,ij}^{A})^2+(\mathbb{Y}_{q, ji}^{A*})^2\right) \cdot B_{2} \cdot \eta_{2}(\mu)\right. \nonumber\\
	&\left.+\ \mathbb{Y}_{q,ij}^{A} \mathbb{Y}_{q,ji}^{A *}\left(\frac{1}{12}+\frac{1}{2} \frac{m_{\phi}^{2}}{\left(m_{i}+m_{j}\right)^{2}}\right) \cdot B_{4} \cdot \eta_{4}(\mu)\right].
	\label{transition element}
\end{align}
This expression gives the observables
\begin{equation}
	\Delta m_{\phi}=2|\Re(\mathcal{M}_{12}^{\phi})| \,,\quad \epsilon_{\phi}=\frac{\Im(\mathcal{M}_{12}^{\phi})}{\sqrt{2}\Delta m_{\phi}}\,.
\end{equation}
For the parameters used in \eqref{transition element}, we follow \cite{Babu:2023oih}.
\begin{itemize}
	\item 
	For the kaon, $B_{2}=0.66$, $B_{4}=1.03$, $f_{K}=160$ MeV, $m_K=0.498$ GeV. 
	The QCD correction factors are $\eta_{2}(\mu)=2.54$ and $\eta_{4}(\mu)=4.81$. 
	The corresponding experimental values for $K^0-\overline{K^0}$ mixing, which limit the New Physics
	contributions, are 
	$\Delta m_{K} = 3.484\times10^{-15}$ GeV and $|\epsilon_{K}| = 2.23\times10^{-3}$.
	\item
	For $B_{d}$, $B_{2}=0.82$, $B_{4}=1.16$, $f_{B_d}=240$ MeV, $m_{B_{d}}=5.28$ GeV. 
	The QCD correction factors are $\eta_{2}(\mu)=2.00$ and $\eta_{4}(\mu)=3.12$. 
	The corresponding experimental value for $B_{d}^{0}-\overline{B_{d}^{0}}$ splitting is 
	$\Delta m_{B_{d}} = 3.12\times10^{-13}$ GeV.
	\item
	For $B_{s}$, $B_{2}=0.82$, $B_{4}=1.16$, $f_{B_s}=295$ MeV, $m_{B_{d}}=5.37$ GeV. 
	The QCD correction factors are $\eta_{2}(\mu)=2.00$ and $\eta_{4}(\mu)=3.12$. 
	The corresponding experimental value for mass splitting is $\Delta m_{B_{s}} = 1.17\times10^{-11}$ GeV.
	\item
	For the $D$ meson, $B_{2}=0.82$, $B_{4}=1.08$, $f_{D}=200$ MeV, $m_D=1.86$ GeV. 
	The QCD correction factors are $\eta_{2}(\mu)=2.31$ and $\eta_{4}(\mu)=3.99$. 
	The mass splitting in the $D$ system is $\Delta m_{D} = 6.25\times10^{-15}$ GeV.
\end{itemize}
In a full calculation, these New Physics contributions should be added to the SM contributions,
and then the total values should be compared with the experimentally observed 
$\Delta m_{\phi}$ and $\epsilon_{\phi}$. However, the SM contributions suffer from uncertainties
stemming from the long-distance hadronic effects. Thus, when checking whether a model
agrees or not with neutral meson oscillation properties, 
we take a poor man's approach: we just require that the New Physics contribution 
does not exceed by absolute value the measured parameters $\Delta m_{\phi}$ and $\epsilon_{\phi}$.

Note that more restrictive prescriptions are sometimes used in literature, 
see e.g. \cite{Ferreira:2019aps}, such as limiting 
the New Physics contribution to $\epsilon_{\phi}$ to 10\% of its experimental value.
We do not expect that this procedure would significantly change our results.
Indeed, the case $(B_1,B_1)$ is already ruled out with our less stringent approach, see Fig.~\ref{fig-B1B1-dead}, 
while the case $(A,B_2)$ can pass the constraint by an order of magnitude or more, Fig.~\ref{fig-D-meson}.

\section{Benchmark points}\label{appendix-benchmark}

The two benchmark points mentioned in the main text correspond to the following 
choice of the parameters in the starting lagrangian.

Benchmark point 1, ``one charged'' boson scenario. The scalar sector parameters are:
\begin{eqnarray}
	&&  m_{11}^2 = 16009~\mbox{GeV}^2\,, \ m_{22}^2 = 6505~\mbox{GeV}^2\,,
	\quad \lambda_1 = 1.9209\,,\quad \lambda_2 = 1.0618\,,\nonumber\\
	&&\lambda_{34} = 0.1857\,,\  \lambda_4 = -0.1215\,,\   
	\lambda_{34}' = -0.09058\,,\ \lambda_4' = -1.1826\,,\ \lambda_5 = -0.6375\,,\nonumber\\
	&& 	\lambda_8 = -0.4785-0.1202\, i\,,\quad \lambda_9 = 0.4846 -0.4713\, i\,.\label{BP1-scalar}
\end{eqnarray}
This potential leads to $v = 246$~GeV, $\tan\beta = 1.8565$ ($\beta = 1.0767$), 
$\tan\psi = 1.2871$ ($\psi = 0.9103$), the angles of the $\hsm$ in the Higgs basis are
\begin{equation}
	\epsilon = -0.2069\,, \quad \alpha = 3.0949\,, \quad \gamma_1 = 0.06963\,, \quad \gamma_2 = 3.3465\,,
\end{equation}
and the scalar masses
\begin{equation}
	m_{H^+_a}\,[\mbox{GeV}] = 149.1,\ 248.4\,, 
	\quad m_{H_A}\,[\mbox{GeV}] = 125.0,\ 186.4,\ 209.6,\ 235.5,\ 324.8\,.
\end{equation}
The parameters of the Yukawa matrices in Eqs.~\eqref{caseA}, \eqref{caseB2} are:
\begin{eqnarray}
	&& g_{11} =0.011758 -0.0035287 i \,, \ g_{12} =0.011496 +0.0038612 i \,, \ g_{13} = 0.0051894 +0.00009072 i\,, \nonumber\\
	&& g_{31} =-0.015744+0.0043714 i \,, \ g_{33} =-0.006668\,,\nonumber\\
	&& d_{11} =-0.17150+0.438410 i \,, \ d_{12} =0.12210 +0.34822 i \,, \ d_{13} =-0.0036445-0.00002292 i \,, \nonumber\\
	&& d_{21} =-0.15711+0.44829 i \,, \ d_{22} =0.13327 +0.34056 i \,, \nonumber\\
	&& d_{31} =0.20164 -0.54406 i \,, \ d_{32} =-0.15670-0.42262 i \,, \ d_{33} =-0.0054627 \,.
	\label{BP1-Yukawa}
\end{eqnarray}
%The quark masses and the CKM matrix are equal to their experimental values used as input. 
The down-quark sector is completely FCNC free at the tree level, while 
the FCNC matrices in the up sector are
\begin{eqnarray}
	N_{u2} &=& \mmmatrix{6.84\times10^{-4}}{8.21\times10^{-7}\,e^{1.49i}}{1.23\times10^{-10}\,e^{-1.65i}}
	{4.00\times10^{-4}\,e^{-1.49i}}{-1.15}{2.23\times10^{-4}}{1.67\times10^{-5}\,e^{1.65i}}
	{6.18\times10^{-2}}{{\bm{92.5}}}\,,\nonumber\\[2mm]
	N_{u3} &=& \mmmatrix{1.33\,e^{-2.22i}}{3.57\times10^{-4}\,e^{-0.731i}}
	{1.44\times10^{-3}\,e^{2.49i}}{1.19\times10^{-4}\,e^{1.69i}}{5.58\times10^{-8}\,e^{-2.73i}}
	{1.89\times10^{-4}\,e^{-2.30i}}{\bm{195.0}\,e^{-0.653i}}{5.26\times10^{-2}\,e^{0.839i}}
	{9.81\times10^{-6}\,e^{0.918i}}\,,
\end{eqnarray}
where for the reader's convenience we highlighted the dominant entries in bold.

Benchmark point 2, ``one neutral'' boson scenario. The scalar sector parameters are:
\begin{eqnarray}
	&&  m_{11}^2 = 2634~\mbox{GeV}^2\,, \ m_{22}^2 = 10109~\mbox{GeV}^2\,,
	\quad \lambda_1 =1.6364 \,,\quad \lambda_2 =0.5297\,,\nonumber\\
	&&\lambda_{34} =-0.4015 \,,\  \lambda_4 =-2.475\,,\   
	\lambda_{34}' =2.1143 \,,\ \lambda_4' =-0.3273\,,\ \lambda_5 =-0.5999\,,\nonumber\\
	&& 	\lambda_8 =-0.7943-0.006037\, i\,,\quad \lambda_9 =0.03356 -0.08534 \, i\,.\label{BP2-scalar}
\end{eqnarray}
This potential leads to $v = 246$~GeV, $\tan\beta = 1.712$ ($\beta = 1.042$), $\tan\psi = 1.073$ ($\psi = 3.962$),
the angles of the $\hsm$ in the Higgs basis are
\begin{equation}
	\epsilon = 0.1238 \,, \quad \alpha =-0.01531 \,, \quad \gamma_1 =3.3657 \,, \quad \gamma_2 = 0.06268\,,
\end{equation}
and the scalar masses
\begin{equation}
	m_{H^+_a}\,[\mbox{GeV}] = 258.9,\ 305.0\,, 
	\quad m_{H_A}\,[\mbox{GeV}] = 125.0,\ 145.7,\ 190.2,\ 265.3,\ 268.7\,.
\end{equation}
The parameters of the Yukawa matrices in Eqs.~\eqref{caseA}, \eqref{caseB2} are:
\begin{eqnarray}
	&& g_{11} =-0.0088293-0.0035565 i \,, \ g_{12} =-0.0027303-0.009088 i \,, \ g_{13} =0.0006978 +0.0006789 i\,, \nonumber\\
	&& g_{31} =0.017254 -0.0077252 i \,, \ g_{33} =-0.00272932\,,\nonumber\\
	&& d_{11} =0.33900 +0.12085 i \,, \ d_{12} =-0.087400-0.32759 i \,, \ d_{13} =0.0030984 +0.0026598 i \,, \nonumber\\
	&& d_{21} =0.093793 -0.35160 i \,, \ d_{22} =-0.31585+0.11259 i \,, \nonumber\\
	&& d_{31} =-0.58535+0.30810 i \,, d_{32} = 0.54538 +0.28705 i \,,  \ \ d_{33} =0.0040640  \,.
	\label{BP2-Yukawa}
\end{eqnarray}
%The quark masses and the CKM matrix are equal to their experimental values used as input. 
The FCNC matrices in the up sector are
\begin{eqnarray}
	N_{u2} &=& \mmmatrix{7.42\times10^{-4}}{8.38\times10^{-7}\,e^{-0.113i}}
	{1.27\times10^{-10}\,e^{3.03i}}{4.08\times10^{-4}\,e^{0.113i}}{-1.06}{2.15\times10^{-4}}
	{1.71\times10^{-5}\,e^{-3.03i}}{5.95\times10^{-2}}{\bm{100.29}}\,,\nonumber\\[2mm]
	N_{u3} &=& \mmmatrix{1.35\,e^{-1.01i}}{3.88\times10^{-4}\,e^{-1.12i}}{1.47\times10^{-3}\,e^{-2.58i}}
	{1.21\times10^{-4}\,e^{2.90i}}{6.07\times10^{-8}\,e^{-3.12i}}{2.06\times10^{-4}\,e^{-2.69i}}
	{\bm{198.8}\,e^{0.564i}}{5.71\times10^{-2}\,e^{0.450i}}{9.998\times10^{-6}\,e^{2.13i}}\,.
\end{eqnarray}

%%%%%%%%%%%%%%%%%%%%%%%%%%%%%%%%%%%%%%%%%%%%%%%%%%%%%%%%%%%

\end{document}